\documentclass[reqno,12pt]{amsart}
\usepackage{amsfonts}
\usepackage{amssymb}
\usepackage[pdftex]{graphicx}
\usepackage{caption}
\usepackage{subcaption}
\usepackage{float}
\usepackage{fullpage}
\usepackage{url}

\begin{document}

\newtheorem{thm}{Theorem}
\newtheorem{lem}{Lemma}
\newtheorem{prop}{Proposition}
\newtheorem{cor}{Corollary}
\numberwithin{equation}{section}



\def\Eqref#1{Eq.~(\ref{#1})}
\def\Eqrefs#1#2{Eqs.~(\ref{#1}) and~(\ref{#2})}
\def\Eqsref#1#2{Eqs.~(\ref{#1}) to~(\ref{#2})}
\def\Sysref#1#2{Eqs. (\ref{#1})--(\ref{#2})}

\def\figref#1{Fig.~\ref{#1}}

\def\secref#1{Sec.~\ref{#1}}
\def\secrefs#1#2{Sec.~\ref{#1} and~\ref{#2}}

\def\appref#1{Appendix~\ref{#1}}

\def\Ref#1{Ref.~\cite{#1}}
\def\Refs#1{Refs.~\cite{#1}}

\def\Cite#1{${\mathstrut}^{\cite{#1}}$}

\def\EQ{\begin{equation}}
\def\endEQ{\end{equation}}


\def\fewquad{\qquad\qquad}
\def\severalquad{\qquad\fewquad}
\def\manyquad{\qquad\severalquad}
\def\manymanyquad{\manyquad\manyquad}

\def\sech{\;{\rm sech}}
\def\sgn{{\rm sgn}}
\def\Re{{\rm Re}\,}
\def\Im{{\rm Im}\,}
\def\cubrt#1{\textstyle\sqrt[3]{#1}}

\def\k{\mathrm{k}}
\def\w{\mathrm{w}}

\def\wtil{\widetilde}

\def\ie/{i.e.}
\def\eg/{e.g.}
\def\etc/{etc.}
\def\const{{\rm const.}}

\allowdisplaybreaks[4]


\title{Oscillatory solitons of U(1)-invariant mKdV equations II: Asymptotic behavior and constants of motion}

\author{
Stephen C. Anco$^1$,
Abdus Sattar Mia$^{1,2}$,
Mark R. Willoughby$^{1}$
\\\\\lowercase{\scshape{
${}^1$
Department of Mathematics\\
Brock University\\
St. Catharines, ON Canada \\
${}^2$
Department of Mathematics and Statistics\\
University of Saskatchewan\\
Saskatoon, SK Canada \\
}}
}

\begin{abstract}
The Hirota equation and the Sasa-Satsuma equation are 
$U(1)$-invariant integrable generalizations of 
the modified Korteweg-de Vries equation. 
These two generalizations admit oscillatory solitons,
which describe harmonically modulated complex solitary waves
parameterized by their speed, modulation frequency, and phase. 
Depending on the modulation frequency, 
the speeds of oscillatory waves ($1$-solitons) 
can be positive, negative, or zero, 
in contrast to the strictly positive speed of ordinary solitons. 
When the speed is zero, an oscillatory wave is a time-periodic standing wave. 
Oscillatory $2$-solitons with non-zero wave speeds are shown to describe 
overtake collisions of a fast wave and a slow wave moving in the same direction,
or head-on collisions of two waves moving in opposite directions. 
When one wave speed is zero, oscillatory $2$-solitons are shown to describe 
collisions in which a moving wave overtakes a standing wave. 
An asymptotic analysis using moving coordinates is carried out to show that, 
in all collisions, 
the speeds and modulation frequencies of the individual waves are preserved, 
while the phases and positions undergo a shift such that 
the center of momentum of the two waves moves at a constant speed. 
The primary constants of motion as well as some other features of 
the nonlinear interaction of the colliding waves are discussed. 
\end{abstract}

\subjclass[2010]{}

\maketitle

\section{ Introduction }

Complex $U(1)$-invariant modified Korteweg-de Vries (mKdV) equations
\begin{equation}
u_t +( \alpha u\bar u_x + \beta u_x\bar u )u + \gamma u_{xxx} =0
\label{genmkdveq}
\end{equation}
(where $\alpha,\beta,\gamma$ are real constants)
arise in many physical applications, 
such as 
short wave pulses in optical fibers 
\cite{Pot,CavCreCru}
and deep water waves
\cite{Sed,Slu}. 
Of particular mathematical and physical interest are 
the two integrable equations in this class, 
given by the Hirota equation \cite{Hir1973}
\begin{equation}
u_t +\beta |u|^2 u_x +\gamma u_{xxx} =0 , 
\end{equation}
and the Sasa-Satsuma equation \cite{SasSat1991}
\begin{equation}
u_t +\alpha( u\bar u_x +3u_x\bar u )u +\gamma u_{xxx} =0 . 
\end{equation}
The integrability properties of these two equations consist of 
multi-soliton solutions, 
a Lax pair, a bilinear formulation, 
a bi-Hamiltonian structure, 
and an infinite hierarchy of symmetries and conservation laws. 

Both equations possess ordinary soliton solutions of the form 
\begin{equation}
u(t,x) = \exp(i\phi) f(x-ct)
\label{solitarywave}
\end{equation}
which are solitary waves 
with speed $c>0$ and phase angle $-\pi\leq \phi\leq \pi$. 
Collisions of two or more solitary waves are described by multi-soliton solutions. 
In all collisions, 
the net effect on the solitary waves is to shift in their asymptotic positions, 
while their asymptotic phase angles 
stay unchanged in the case of the Hirota equation \eqref{hmkdveq}
but become shifted in the case of the Sasa-Satsuma equation \eqref{ssmkdveq}. 
The actual nonlinear interaction of these solitary waves during a collision 
exhibits interesting features which depend on the speed ratios 
and relative phase angles of the waves,
as studied in previous work \cite{AncNgaWil}. 
(See the animations at 
http://lie.math.brocku.ca/\url{~}sanco/solitons/mkdv\_solitons.php)

In a recent paper \cite{AncMiaWil}, 
we began a comprehensive study of a more general type of soliton solution 
with the form 
\begin{equation}
u(t,x) = \exp(i\phi)\exp(i\nu t) \tilde f(x-ct)
\label{oscil1soliton}
\end{equation}
which is a harmonically modulated solitary wave, 
called an {\em oscillatory soliton}, 
where the temporal modulation frequency $\nu\neq 0$ and the speed $c$
obey the kinematic relation 
\begin{equation}\label{kincond}
(c/3)^3 +(\nu/2)^2 >0 . 
\end{equation}
This relation is required for the oscillatory soliton to be a solution of
the the Hirota equation \eqref{hmkdveq}
and the Sasa-Satsuma equation \eqref{ssmkdveq}. 
In contrast to an ordinary soliton, 
the speed $c$ of an oscillatory soliton can be positive, negative, or zero. 
Consequently, 
these solitons have three different types of collisions:
(1) {\em right-overtake} --- 
where a faster right-moving soliton overtakes a slower right-moving soliton 
or a stationary soliton;
(2) {\em left-overtake} --- 
where a faster left-moving soliton overtakes a slower left-moving soliton
or a stationary soliton;
(3) {\em head-on} --- 
where a right-moving soliton collides with a left-moving soliton. 
All of these collisions are described by oscillatory $2$-soliton solutions
\begin{equation}
\begin{aligned}
u(t,x) =& \exp(i\phi_1)\exp(i\nu_1 t) \tilde f_1(x-c_1 t,x-c_2 t) 
\\&\qquad
+ \exp(i\phi_2)\exp(i\nu_2 t) \tilde f_2(x-c_1 t,x-c_2 t), 
\quad
c_1\neq c_2
\end{aligned}
\label{oscil2soliton}
\end{equation}
whose temporal frequencies $\nu_1$, $\nu_2$ and speeds $c_1$, $c_2$ 
satisfy the kinematic relations
\begin{equation}\label{kinconds}
(c_1/3)^3 +(\nu_1/2)^2 >0 ,
\quad
(c_2/3)^3 +(\nu_2/2)^2 >0 . 
\end{equation}

In the present paper 
we will study the asymptotic features of 
colliding oscillatory solitons \eqref{oscil2soliton}
for both the Hirota equation \eqref{hmkdveq}
and the Sasa-Satsuma equation \eqref{ssmkdveq}. 
These collisions can be expected to exhibit 
highly interesting new features compared to collisions of ordinary solitons.

In section~\ref{solns}, 
we recall some details of the oscillatory $1$-soliton and $2$-soliton solutions
derived in \Ref{AncMiaWil} 
for the Hirota equation and the Sasa-Satsuma equation. 

In section~\ref{asymptotics},
we carry out an asymptotic analysis using moving coordinates 
to show that the $2$-soliton solutions \eqref{oscil2soliton}
for the Hirota equation and the Sasa-Satsuma equation 
reduce to a superposition of two $1$-soliton solutions \eqref{oscil1soliton}
with speeds $c_1$, $c_2$ and temporal frequencies $\nu_1$, $\nu_2$
in the asymptotic past and future. 
This analysis rigorously establishes that these $2$-soliton solutions 
describe collisions between two oscillatory waves
when both $c_1$ and $c_2$ are non-zero, 
or collisions of an oscillatory wave with a standing wave, 
when one of the speeds $c_1$ or $c_2$ is zero. 

In section~\ref{com},
for the $1$-soliton and $2$-soliton solutions, 
we discuss the primary constants of motion 
arising from the conservation laws for momentum, energy, and Galilean energy 
admitted by \cite{conslaws} the Hirota equation and the Sasa-Satsuma equation. 
In particular, from conservation of Galilean energy, 
we show that the center of momentum for the $2$-soliton solutions 
moves at a constant speed throughout a collision. 

Our main results are obtained in section~\ref{results}.
For overtake and head-on collisions described by the $2$-soliton solutions, 
we first show that the net effect of a collision is to shift 
the asymptotic positions and phases of the two waves 
while the speed and the temporal frequency of each wave remains unchanged. 
Explicit formulas for these asymptotic shifts are presented 
in terms of the speeds and temporal frequencies of the two waves
in the collision. 
Next, from these formulas, 
we find that for overtake collisions 
the faster wave gets shifted forward relative to its direction of motion 
while the slower or stationary wave gets shifted in the backward direction. 
In contrast, for head-on collisions, 
we find that both waves get shifted forward 
relative to their directions of motion. 
Finally, for all collisions, 
we show that the position shifts of the two oscillatory waves
are related by the property that the center of momentum of the waves
is preserved in the collision. 

In section~\ref{interactions}, 
we discuss a few interesting features of the nonlinear interactions that occur 
for oscillatory waves and standing waves during collisions, 
such as the appearance of nodes, phase coils, and phase reversals. 
We also make some concluding remarks. 

Previous work on soliton solutions of 
the Hirota equation and Sasa-Satsuma equation
appears in \Ref{Mih1993,Mih1994,TaoHe,GilHieNimOht}. 
This work amounts to deriving the $1$-soliton and $2$-soliton formulas
in a mathematically equivalent but less physically useful envelope form,
without any analysis of the asymptotic behaviour and the constants of motion
for these solutions. 

All computations in the present paper have been carried out by use of Maple. 
Hereafter, by scaling variables $t,x,u$, we will put 
\EQ\label{scaling}
\alpha=6, 
\quad
\beta=24, 
\quad
\gamma=1
\endEQ 
for convenience.

\section{Oscillatory soliton solutions}
\label{solns} 

For the Hirota equation 
\begin{equation}
u_t +24 |u|^2 u_x + u_{xxx} =0
\label{hmkdveq}
\end{equation}
and the Sasa-Satsuma equation 
\begin{equation}
u_t +6( u\bar u_x +3u_x\bar u )u +\gamma u_{xxx} =0
\label{ssmkdveq}
\end{equation}
we will first summarize the expressions 
for the respective travelling-wave functions $\tilde f(x-ct)$
in the oscillatory $1$-soliton solutions \eqref{oscil1soliton}. 

Let 
\begin{gather}
k = \frac{\sqrt{3}}{2}\Big( \cubrt{\sqrt{(c/3)^3 +(\nu/2)^2} - \nu/2} 
+ \cubrt{\sqrt{(c/3)^3 +(\nu/2)^2} + \nu/2}\; \Big) , 
\label{krel}
\\
\kappa = \frac{1}{2}\Big( \cubrt{\sqrt{(c/3)^3 +(\nu/2)^2} - \nu/2} 
- \cubrt{\sqrt{(c/3)^3 +(\nu/2)^2} + \nu/2}\; \Big) , 
\label{kapparel}
\end{gather}
where $c$ and $\nu\neq0$ obey the relation \eqref{kincond},
which corresponds to the properties $k>0$ and $\kappa\neq 0$. 

\begin{prop}\label{thm:1soliton}
The Hirota and Sasa-Satsuma 
oscillatory $1$-soliton solutions 
\begin{equation}\label{1soliton}
u(t,x) = \exp(i\phi)\exp(i\nu t) \tilde f(\xi),
\qquad
\xi=x-ct
\end{equation}
expressed using a travelling wave coordinate are given by 
\begin{equation}\label{f1soliton}
\tilde f(\xi) = \exp(i\kappa\xi) U(\xi)
\end{equation}
in terms of the respective envelope functions
\begin{gather}
U_{\rm H}(\xi) = 
\frac{k}{2\cosh(k\xi)} , 
\label{hoscilfunct}
\\
U_{\rm SS}(\xi) = 
\frac{k(2|\kappa|)^{1/2}(k^2+\kappa^2)^{1/4}\cosh(k\xi +i\lambda/2)}{|\kappa|\cosh(2k\xi) +(k^2+\kappa^2)^{1/2}} ,
\quad
\lambda = \arg(\kappa(\kappa+ik)) ,
\label{ssoscilfunct}
\end{gather}
with $\arg(U_{\rm H}) =0$ and $\arg(U_{\rm SS}) =\arctan(\tanh(k\xi)\tan(\lambda/2))$.
When $c=0$, 
these $1$-soliton solutions are harmonically modulated standing waves. 
\end{prop}

We next summarize the expressions for the travelling-wave functions
$\tilde f_1(x-c_1 t,x-c_2 t)$ and $\tilde f_2(x-c_1 t,x-c_2 t)$
in oscillatory $2$-soliton solutions \eqref{oscil2soliton}
for the Hirota equation \eqref{hmkdveq}
and the Sasa-Satsuma equation \eqref{ssmkdveq}. 

Let 
\begin{gather}
k_1 = \sqrt{3}(\beta_{1-} + \beta_{1+})/2 ,
\quad
\kappa_1 = (\beta_{1-} - \beta_{1+})/2 ,
\label{rels1}
\\
k_2 = \sqrt{3}(\beta_{2-} + \beta_{2+})/2 ,
\quad
\kappa_2 = (\beta_{2-} - \beta_{2+})/2 ,
\label{rels2}
\end{gather}
with 
\begin{equation}\label{betapms}
\beta_{1\pm} = \cubrt{\sqrt{(c_1/3)^3 +(\nu_1/2)^2} \pm \nu_1/2} ,
\quad
\beta_{2\pm} = \cubrt{\sqrt{(c_2/3)^3 +(\nu_2/2)^2} \pm \nu_2/2} , 
\end{equation}
where $c_1$, $c_2$, $\nu_1\neq0$, $\nu_2\neq 0$
obey the relations \eqref{kinconds},
corresponding to the properties 
\begin{gather}
k_1>0 ,
\quad
k_2>0 , 
\label{k1k2prop}
\\
\kappa_1\neq0 ,
\quad
\kappa_2\neq 0 . 
\end{gather}

\begin{prop}\label{thm:2soliton}
As expressed using travelling wave coordinates 
$\xi_1=x-c_1 t$ and $\xi_2=x-c_2 t$ when $c_1\neq c_2$, 
the Hirota and Sasa-Satsuma
oscillatory $2$-soliton solutions 
\begin{equation}\label{2soliton}
u(t,x) = \exp(i\phi_1)\exp(i\nu_1 t) \tilde f_1(\xi_1,\xi_2) + \exp(i\phi_2)\exp(i\nu_2 t) \tilde f_2(\xi_1,\xi_2)
\end{equation}
are given by 
\begin{equation}\label{f2soliton}
\tilde f_1(\xi_1,\xi_2) = \exp(i\kappa_1\xi_1) V_1(\xi_1,\xi_2)/W(\xi_1,\xi_2),
\quad
\tilde f_2(\xi_1,\xi_2) =\exp(i\kappa_2\xi_2) V_2(\xi_1,\xi_2)/W(\xi_1,\xi_2)
\end{equation}
in terms of the respective envelope functions
\begin{align}
& V_{1\rm H}(\xi_1,\xi_2)
= k_1 \cosh(k_2\xi_2 +i\gamma_2) , 
\label{hoscilX1}\\
& V_{2\rm H}(\xi_1,\xi_2)
= k_2 \cosh(k_1\xi_1 +i\gamma_1) , 
\label{hoscilX2}\\
&\begin{aligned}
W_{\rm H}(\xi_1,\xi_2)
= & \sqrt{\Gamma}\cosh(k_1\xi_1+k_2\xi_2) +\frac{1}{\sqrt{\Gamma}}\cosh(k_1\xi_1-k_2\xi_2) 
\\&\qquad
-\frac{4k_1k_2}{\sqrt{\Upsilon}} \cos(\kappa_1\xi_1-\kappa_2\xi_2+\mu(\xi_2-\xi_1)+\phi_1-\phi_2) , 
\end{aligned}
\label{hoscilY}
\end{align}
in the Hirota case, 
and 
\begin{align}
&\begin{aligned}
V_{1\rm SS}(\xi_1,\xi_2) = 
& 
k_1(k_1^2+\kappa_1^2)^{1/4} |2\kappa_1|^{1/2}
\bigg(
|\kappa_2|\Big( \sqrt{\Delta\Gamma} \cosh(k_1\xi_1+2k_2\xi_2+i(\alpha_2+\gamma_2))
\\&
+\frac{1}{\sqrt{\Delta\Gamma}}\cosh(k_1\xi_1-2k_2\xi_2+i(\upsilon_2-\gamma_2)) \Big)
\\&
+(k_2^2+\kappa_2^2)^{1/2}\Big( 
-8k_2^2\kappa_2 s \frac{1}{\sqrt{\Omega\Upsilon}} \cosh(k_1\xi_1+i(\varpi_1+\gamma_1))
\\&
+\sqrt{\frac{\Gamma}{\Delta}} \cosh(k_1\xi_1+i(\alpha_2-\gamma_2))
+ \sqrt{\frac{\Delta}{\Gamma}} \cosh(k_1\xi_1+i(\upsilon_2+\gamma_2)) \Big)
\bigg)
\\&
+ k_1k_2 s \sqrt{\frac{\Omega}{\Upsilon}} \bigg(
k_2 (k_1^2+\kappa_1^2)^{1/4} |8\kappa_2|^{1/2} s_2 \cosh(k_1\xi_1+i(\varpi_1-\gamma_1)) 
\\&\qquad
- k_1(k_2^2+\kappa_2^2)^{1/4} |32\kappa_1|^{1/2} s_1
\Re\Big( \cosh(k_2\xi_2+i(\gamma_2-\varpi_2)) 
\times
\\&\qquad\qquad
\exp(i(\kappa_1\xi_1-\kappa_2\xi_2+\mu(\xi_2-\xi_1)+\phi_1-\phi_2)) \Big) \bigg) ,
\end{aligned}
\label{ssoscilX1}\\
&\nonumber\\
&\nonumber\\
&\begin{aligned}
V_{2\rm SS}(\xi_1,\xi_2) = 
& 
k_2(k_2^2+\kappa_2^2)^{1/4} |2\kappa_2|^{1/2}
\bigg(
|\kappa_1|\Big( \sqrt{\Delta\Gamma} \cosh(k_2\xi_2+2k_1\xi_1+i(\alpha_1+\gamma_1))
\\& 
+\frac{1}{\sqrt{\Delta\Gamma}}\cosh(k_2\xi_2-2k_1\xi_1+i(\upsilon_1-\gamma_1)) \Big)
\\&
+(k_1^2+\kappa_1^2)^{1/2}\Big( 
-8k_1^2\kappa_1 s \frac{1}{\sqrt{\Omega\Upsilon}} 
\cosh(k_2\xi_2+i(\varpi_2+\gamma_2)) 
\\&
+\sqrt{\frac{\Gamma}{\Delta}} \cosh(k_2\xi_2+i(\alpha_1-\gamma_1))
+ \sqrt{\frac{\Delta}{\Gamma}} \cosh(k_2\xi_2+i(\upsilon_1+\gamma_1)) \Big)
\bigg)
\\&
+ k_1k_2 s \sqrt{\frac{\Omega}{\Upsilon}} \bigg(
k_1(k_2^2+\kappa_2^2)^{1/4} |8\kappa_1|^{1/2} s_1 \cosh(k_2\xi_2+i(\varpi_2-\gamma_2)) 
\\&\qquad
- k_2(k_1^2+\kappa_1^2)^{1/4} |32\kappa_2|^{1/2} s_2
\Re\Big( \cosh(k_1\xi_1+i(\gamma_1-\varpi_1)) 
\times
\\&\qquad\qquad
\exp(i(\kappa_2\xi_2-\kappa_1\xi_1+\mu(\xi_1-\xi_2)+\phi_2-\phi_1)) \Big) \bigg) ,
\end{aligned}
\label{ssoscilX2}\\
&\nonumber\\
&\nonumber\\
&\begin{aligned}
W_{\rm SS}(\xi_1,\xi_2)= &
|\kappa_1\kappa_2|\Big( \Delta\Gamma \cosh(2(k_1\xi_1+k_2\xi_2))
+ \frac{1}{\Delta\Gamma} \cosh(2(k_1\xi_1-k_2\xi_2)) \Big)
\\&
+2(k_1^2+\kappa_1^2)^{1/2}|\kappa_2| \cosh(2k_2\xi_2) 
+ 2(k_2^2+\kappa_2^2)^{1/2}|\kappa_1| \cosh(2k_1\xi_1)
\\&
+4k_1^2 k_2^2s_1s_2\frac{\Omega}{\Upsilon}
\cos(2(\kappa_1\xi_1-\kappa_2\xi_2+\mu(\xi_2-\xi_1))+2(\phi_1-\phi_2))
\\&
+(k_1^2+\kappa_1^2)^{1/2}(k_2^2+\kappa_2^2)^{1/2}
\bigg(\frac{\Gamma}{\Delta} +\frac{\Delta}{\Gamma} 
+64k_1^2 k_2^2\kappa_1\kappa_2 \frac{1}{\Omega\Upsilon}
\\&\qquad
-16 k_1k_2 |\kappa_1\kappa_2|^{1/2} 
\Re\Big( \exp(i(\kappa_1\xi_1-\kappa_2\xi_2+\mu(\xi_2-\xi_1)+\phi_1-\phi_2))
\times
\\&\qquad\qquad
\Big( \sqrt{\frac{\Gamma}{\Upsilon}} \cosh(k_1\xi_1+k_2\xi_2+i(\varpi_1-\varpi_2))
\\&\qquad\qquad\qquad
+ \frac{1}{\sqrt{\Gamma\Upsilon}} \cosh(k_1\xi_1-k_2\xi_2+i(\varpi_1+\varpi_2)) \Big) \Big)
\bigg) 
\end{aligned}
\label{ssoscilY}
\end{align}
in the Sasa-Satsuma case. 
In both cases, 
\begin{gather}
\mu = (\nu_1-\nu_2)/(c_1-c_2) ,
\label{2solitonmu}
\\
\Omega = \sqrt{\big((k_1+k_2)^2+(\kappa_1+\kappa_2)^2\big)\big((k_1 -k_2)^2+(\kappa_1 +\kappa_2)^2\big)} , 
\label{2solitonOmega}
\\
\Upsilon = \big((k_1-k_2)^2+(\kappa_1 -\kappa_2)^2\big)\big((k_1+k_2)^2+(\kappa_1-\kappa_2)^2\big) ,
\label{2solitonUpsilon}
\\
\Delta = \sqrt{\frac{(k_1-k_2)^2+(\kappa_1+\kappa_2)^2}{(k_1 +k_2)^2+(\kappa_1 +\kappa_2)^2}} ,
\label{2solitonDelta}
\\
\Gamma = \frac{(k_1-k_2)^2+(\kappa_1-\kappa_2)^2}{(k_1 +k_2)^2+(\kappa_1 -\kappa_2)^2} ,
\label{2solitonGamma}
\\
\alpha_1 = (\lambda_2+\delta_1)/2 , 
\quad
\alpha_2 = (\lambda_1+\delta_2)/2 ,
\label{2solitonalphas}
\\
\upsilon_1 = (\lambda_2-\delta_1)/2 , 
\quad
\upsilon_2 = (\lambda_1-\delta_2)/2 ,
\label{2solitonupsilons}
\\
\varpi_1=(\lambda_1 -\delta_1)/2 , 
\quad
\varpi_2=(\lambda_2 -\delta_2)/2 ,
\label{2solitonvarpis}
\\
\begin{aligned}
\gamma_1=\arg( k_1^2-k_2^2 -(\kappa_1-\kappa_2)^2 +i2k_1(\kappa_1-\kappa_2) ) ,
\\
\gamma_2=\arg( k_2^2-k_1^2 -(\kappa_1-\kappa_2)^2 -i2k_2(\kappa_1-\kappa_2) ) ,
\end{aligned}
\label{2solitongammas}
\\
\begin{aligned}
\delta_1=\arg( k_2^2-k_1^2 +(\kappa_1+\kappa_2)^2 +i2k_1(\kappa_1+\kappa_2) ) ,
\\
\delta_2=\arg( k_1^2-k_2^2 +(\kappa_1+\kappa_2)^2+i2k_2(\kappa_1+\kappa_2) ) ,
\end{aligned}
\label{2solitondeltas}
\\
\lambda_1 = \arg( \kappa_1(\kappa_1+ik_1) ) ,
\quad
\lambda_2 = \arg( \kappa_2(\kappa_2+ik_2) ) .
\label{2solitonlambdas}
\\
s_1=\sgn(\kappa_1) ,
\quad
s_2=\sgn(\kappa_2) ,
\label{sgns}
\\
s=\begin{cases}
1, & |k_1|\neq|k_2|\\
\sgn(\kappa_1+\kappa_2), & |k_1|=|\k_2|
\end{cases} .
\label{sssgn}
\end{gather}
\end{prop}

For convenience, 
we also write out the half-angle expressions 
needed in equations \eqref{2solitonalphas}--\eqref{2solitonvarpis}:
\begin{align}
\lambda_1/2 = \arg\Big(\sqrt{\sqrt{1+k_1^2/\kappa_1^2} +1} +is_1\sqrt{\sqrt{1+k_1^2/\kappa_1^2} -1}\Big) ,
\label{2solitonhalflambda1}
\\
\lambda_2/2 = \arg\Big(\sqrt{\sqrt{1+k_2^2/\kappa_2^2} +1} +is_2\sqrt{\sqrt{1+k_2^2/\kappa_2^2} -1}\Big) ,
\label{2solitonhalflambda2}
\end{align}
and
\begin{align}
\begin{aligned}
\delta_1/2=& 
\arg\Big( (\epsilon^2\epsilon_- + (1-\epsilon^2\epsilon_-)\sgn(\kappa_1+\kappa_2))
 \sqrt{k_2^2-k_1^2 +(\kappa_1+\kappa_2)^2+\Omega} 
\\&\qquad
+i s_1( 1+\epsilon^2\epsilon_-(\sgn(\kappa_1+\kappa_2)-1)) \sqrt{k_2^2-k_1^2 +(\kappa_1+\kappa_2)^2-\Omega} \Big) ,
\end{aligned}
\label{2solitonhalfdelta1}
\\
\begin{aligned}
\delta_2/2=&
\arg\Big( (\epsilon^2\epsilon_+ + (1-\epsilon^2\epsilon_+)\sgn(\kappa_1+\kappa_2))\sqrt{k_1^2-k_2^2 +(\kappa_1+\kappa_2)^2+\Omega} 
\\&\qquad 
+i s_2( 1+\epsilon^2\epsilon_+(\sgn(\kappa_1+\kappa_2)-1))\sqrt{k_1^2-k_2^2 +(\kappa_1+\kappa_2)^2-\Omega} \Big) ,
\end{aligned}
\label{2solitonhalfdelta2}
\end{align}
where 
\begin{equation}\label{sgnss}
\epsilon_\pm = (1\pm \epsilon)/2,  
\quad
\epsilon=\sgn(|k_1|-|k_2|) 
=\begin{cases}
1, & |k_1|>|k_2|\\
-1, & |k_1|<|k_2|\\
0, & |k_1|=|k_2|
\end{cases} .
\end{equation}

\section{Asymptotic analysis}
\label{asymptotics} 

An oscillatory wave \eqref{oscil1soliton}
having phase angle $\phi$, temporal frequency $\nu$, and speed $c$ 
reduces to an ordinary travelling wave \eqref{solitarywave}
when (and only when) $\nu =0$. 
In this case the kinematic relation \eqref{kincond} implies $c>0$,
showing that all ordinary travelling wave solutions of 
the Hirota equation \eqref{hmkdveq}
and the Sasa-Satsuma equation \eqref{ssmkdveq} 
are right-moving. 

In contrast, when $\nu\neq0$, 
the kinematic relation \eqref{kincond} states 
\begin{equation}\label{1solitonkinrel}
c > -(3/\cubrt{4})(\cubrt{\nu})^2 \neq 0 
\end{equation}
which allows $c<0$ and $c=0$, in addition to allowing $c>0$. 
Correspondingly, oscillatory $1$-soliton solutions 
from Proposition~\ref{thm:1soliton} for 
the Hirota equation \eqref{hmkdveq}
and the Sasa-Satsuma equation \eqref{ssmkdveq} 
consist of right-moving oscillatory waves when $c>0$, 
left-moving oscillatory waves when $c<0$, 
and standing waves when $c=0$. 

We will now show that the oscillatory $2$-soliton solutions 
with $c_1\neq c_2$ from Proposition~\ref{thm:2soliton}
for the Hirota equation \eqref{hmkdveq}
and the Sasa-Satsuma equation \eqref{ssmkdveq} 
reduce in both the asymptotic past ($t\rightarrow -\infty$) and future ($t\rightarrow +\infty$)
to a linear superposition of oscillatory $1$-soliton solutions
whose speeds are precisely $c_1$ and $c_2$. 
Since the solutions are symmetric under simultaneously interchanging 
$c_1 \longleftrightarrow c_2$, $\nu_1 \longleftrightarrow \nu_2$, 
$\phi_1 \longleftrightarrow \phi_2$, 
we will assume 
\begin{equation}
c_1>c_2
\end{equation}
hereafter without loss of generality. 

To proceed, we first note 
\begin{equation}\label{movingcoords}
\xi_1 = x -c_1 t, 
\quad
\xi_2 = x -c_2 t
\end{equation}
are moving coordinates centered at positions $x=c_1t$ and $x=c_2t$, respectively.
Consider 
\begin{equation}\label{asymptparm}
\varepsilon = \xi_2 - \xi_1 = (c_1-c_2)t 
\end{equation}
with $c_1-c_2>0$ whereby 
$\xi_1$ is the rightmost coordinate and $\xi_2$ is the leftmost coordinate. 

Asymptotic expansions for $t\rightarrow \pm\infty$ 
then correspond to asymptotic expansions given by 
$\varepsilon \rightarrow \pm\infty$. 
In each expansion, 
we separately hold fixed the coordinates $\xi_1$ and $\xi_2$.

\subsection{Moving-coordinate expansion of the Hirota oscillatory $2$-soliton}
We begin by holding the rightmost coordinate fixed, 
and expressing the leftmost coordinate in terms of 
the expansion parameter $\varepsilon$ from equation \eqref{asymptparm}, 
so thus 
\begin{equation}\label{fastexpansion}
\xi_1=\const,
\quad 
\xi_2= \xi_1 +\varepsilon \rightarrow \pm\infty 
\end{equation}
as $\varepsilon \rightarrow \pm\infty$. 
In the Hirota oscillatory $2$-soliton solution from Proposition~\ref{thm:2soliton}, 
applying the expansion \eqref{fastexpansion} to the functions \eqref{hoscilX1}, \eqref{hoscilX2}, \eqref{hoscilY} 
and neglecting subdominant terms, we find 
\begin{align}
& e^{i\kappa_1\xi_1} V_{1\rm H}(\xi_1,\xi_1+\varepsilon) 
\sim e^{k_2|\varepsilon|}\Big( \tfrac{1}{2}k_1 e^{\pm i\gamma_2} e^{i\kappa_1\xi_1} e^{\pm k_2\xi_1} \Big) +O(1)
\label{hX1fast}
\\
& e^{i\kappa_2(\xi_1+\varepsilon)} V_{2\rm H}(\xi_1,\xi_1+\varepsilon)  
\sim O(1)
\label{hX2fast}
\\
& W_{\rm H}(\xi_1,\xi_1+\varepsilon)  
\sim e^{k_2|\varepsilon|}\Big( \tfrac{1}{2}k_1 e^{\pm k_2\xi_1} \Big(\sqrt{\Gamma} e^{\pm k_1\xi_1} +\frac{1}{\sqrt{\Gamma}} e^{\mp k_1\xi_1}\Big)
\Big) +O(1) 
\label{hYfast}
\end{align}
(with $k_2>0$ due to equation \eqref{k1k2prop}). 
Up to the exponential factor $e^{k_2|\varepsilon|}e^{\pm k_2\xi_1}$, 
the asymptotic functions \eqref{hX1fast} and \eqref{hYfast}
resemble the form of the numerator and denominator 
in the oscillatory $1$-soliton solution \eqref{f1soliton} and \eqref{hoscilfunct} 
for the Hirota equation. 
Specifically, 
the expansion \eqref{hYfast} can be expressed as 
\begin{equation}
e^{-k_2|\varepsilon|} e^{\mp k_2\xi_1} W_{\rm H}(\xi_1,\xi_1+\varepsilon)  
\sim \cosh(k_1\xi_1^\pm), 
\qquad \varepsilon\rightarrow \pm\infty
\end{equation}
where
\begin{equation}\label{hshiftcoord1}
\xi_1^\pm = \xi_1 \mp x_1, 
\quad
x_1 = -\frac{\ln\Gamma}{2k_1} >0 
\end{equation}
is a shifted moving coordinate. 
Then the expansion \eqref{hX1fast} can be written in terms of this coordinate,
which gives
\begin{equation}
e^{-k_2|\varepsilon|} e^{\mp k_2\xi_1} e^{i\kappa_1\xi_1} V_{1\rm H}(\xi_1,\xi_1+\varepsilon)  
\sim \tfrac{1}{2}k_1 e^{i(\pm\gamma_2 \pm\kappa_1 x_1)} e^{i\kappa_1\xi_1^\pm}, 
\qquad \varepsilon\rightarrow \pm\infty . 
\end{equation}
Hence we have 
\begin{equation}\label{hf1fast}
e^{i\kappa_1\xi_1} V_{1\rm H}(\xi_1,\xi_1+\varepsilon)/W_{\rm H}(\xi_1,\xi_1+\varepsilon)
= \tilde f_{1\rm H}(\xi_1,\xi_1+\varepsilon) 
\sim e^{i(\pm\gamma_2 \pm\kappa_1 x_1)} e^{i\kappa_1\xi_1^\pm} U_{\rm H}(\xi_1^\pm), 
\qquad \varepsilon\rightarrow \pm\infty
\end{equation}
where $\tilde f_{1\rm H}$ is the function \eqref{f1soliton}
and $U_{\rm H}$ is the function \eqref{hoscilfunct}. 
Finally, 
the phase factor $e^{i(\pm\gamma_2 \pm\kappa_1 x_1)}$ can be combined with $e^{i\phi_1}$ 
to get a shifted phase angle
\begin{equation}\label{hshiftphi1}
\phi_1^\pm= \phi_1 \pm\eta_1 , 
\quad
\eta_1 = \gamma_2 +\kappa_1 x_1 .
\end{equation}
In a similar way, the expansion \eqref{hX2fast} gives
\begin{equation}
e^{-k_2|\varepsilon|} e^{\mp k_2\xi_1} e^{i\kappa_2(\xi_1+\varepsilon)} V_{2\rm H}(\xi_1,\xi_1+\varepsilon)  
\sim 0, 
\qquad \varepsilon\rightarrow \pm\infty
\end{equation}
whence 
\begin{equation}\label{hf2fast}
e^{i\kappa_2(\xi_1+\varepsilon)} V_{2\rm H}(\xi_1,\xi_1+\varepsilon)/W_{\rm H}(\xi_1,\xi_1+\varepsilon)
= \tilde f_{2\rm H}(\xi_1,\xi_1+\varepsilon) 
\sim 0 , 
\qquad \varepsilon\rightarrow \pm\infty .
\end{equation}

Combining equations \eqref{hf1fast} and \eqref{hf2fast} with equation \eqref{f2soliton}, 
we see that the asymptotic expansion of the Hirota oscillatory $2$-soliton solution
with respect to its leftmost coordinate is given by 
\begin{equation}\label{hmovingcoordfast}
u(t,x) \sim e^{i\phi_1^\pm} e^{i\nu_1 t} e^{i\kappa_1 \xi_1^\pm} U_{\rm H}(\xi_1^\pm) = u_1^\pm(t,x), 
\qquad \xi_1=\const,\quad \varepsilon=\xi_2-\xi_1\rightarrow \pm\infty .
\end{equation}

Next, we hold the leftmost coordinate fixed, 
and express the rightmost coordinate in terms of 
the expansion parameter $\varepsilon$ from equation \eqref{asymptparm}, 
so now
\begin{equation}\label{slowexpansion} 
\xi_2=\const,
\quad 
\xi_1= \xi_2 -\varepsilon \rightarrow \mp\infty 
\end{equation}
as $\varepsilon \rightarrow \pm\infty$. 
Applying this expansion \eqref{slowexpansion} to the functions \eqref{hoscilX1}, \eqref{hoscilX2}, \eqref{hoscilY}, 
we obtain 
\begin{align}
& e^{i\kappa_1 (\xi_2-\varepsilon)}V_{1\rm H}(\xi_2-\varepsilon,\xi_2) 
\sim O(1)
\label{hX1slow}
\\
& e^{i\kappa_2\xi_2} V_{2\rm H}(\xi_2-\varepsilon,\xi_2)   
\sim e^{k_1|\varepsilon|}\Big( \tfrac{1}{2}k_2 e^{\mp i\gamma_1} e^{i\kappa_2\xi_2} e^{\mp k_1\xi_2} \Big) +O(1)
\label{hX2slow}
\\
& W_{\rm H}(\xi_2-\varepsilon,\xi_2) 
\sim e^{k_1|\varepsilon|}\Big( \tfrac{1}{2}k_2 e^{\mp k_1\xi_2} \Big(\sqrt{\Gamma} e^{\mp k_2\xi_2} +\frac{1}{\sqrt{\Gamma}} e^{\pm k_2\xi_2}\Big)
\Big) +O(1)
\label{hYslow}
\end{align}
(with $k_1>0$ due to equation \eqref{k1k2prop}). 
The expansions \eqref{hYslow} and \eqref{hX2slow} can be expressed as 
\begin{equation}
e^{-k_1|\varepsilon|} e^{\pm k_1\xi_2} W_{\rm H}(\xi_2-\varepsilon,\xi_2) 
\sim \cosh(k_2\xi_2^\pm), 
\qquad \varepsilon\rightarrow \pm\infty
\end{equation}
and
\begin{equation}
e^{-k_1|\varepsilon|} e^{\pm k_1\xi_2} e^{i\kappa_2\xi_2} V_{2\rm H}(\xi_2-\varepsilon,\xi_2) 
\sim \tfrac{1}{2}k_2 e^{i(\mp\gamma_1 \pm\kappa_2 x_2)} e^{i\kappa_2\xi_2^\pm}, 
\qquad \varepsilon\rightarrow \pm\infty
\end{equation}
where
\begin{equation}\label{hshiftcoord2}
\xi_2^\pm = \xi_2 \mp x_2, 
\quad
x_2 = \frac{\ln\Gamma}{2k_2} <0 
\end{equation}
is a shifted moving coordinate. 
Hence we have 
\begin{equation}\label{hf2slow}
e^{i\kappa_2\xi_2} V_{2\rm H}(\xi_2-\varepsilon,\xi_2)/W_{\rm H}(\xi_2-\varepsilon,\xi_2)
= \tilde f_{2\rm H}(\xi_2-\varepsilon,\xi_2) 
\sim e^{i(\mp\gamma_1 \pm\kappa_2 x_2)} e^{i\kappa_2\xi_2^\pm} U_{\rm H}(\xi_2^\pm) , 
\qquad \varepsilon\rightarrow \pm\infty .
\end{equation}
The phase factor $e^{i(\mp\gamma_1 \pm\kappa_2 x_2)}$ can be combined with $e^{i\phi_2}$ 
to get a shifted phase angle
\begin{equation}\label{hshiftphi2}
\phi_2^\pm= \phi_2 \pm\eta_2 , 
\quad
\eta_2 = -\gamma_1 +\kappa_2 x_2 .
\end{equation}
Similarly, the expansion \eqref{hX1slow} gives
\begin{equation}
e^{-k_1|\varepsilon|} e^{\pm k_1\xi_2} e^{i\kappa_1(\xi_2-\varepsilon)}V_{1\rm H}(\xi_2-\varepsilon,\xi_2) 
\sim 0, 
\qquad \varepsilon\rightarrow \pm\infty
\end{equation}
whence 
\begin{equation}\label{hf1slow}
e^{i\kappa_1(\xi_2-\varepsilon)} V_{1\rm H}(\xi_2-\varepsilon,\xi_2)/W_{\rm H}(\xi_2-\varepsilon,\xi_2)
= \tilde f_{1\rm H}(\xi_2-\varepsilon,\xi_2) 
\sim 0 , 
\qquad \varepsilon\rightarrow \pm\infty .
\end{equation}

Combining equations \eqref{hf1slow} and \eqref{hf2slow} with equation \eqref{f2soliton}, 
we see that the asymptotic expansion of the Hirota oscillatory $2$-soliton solution
with respect to its rightmost coordinate is given by 
\begin{equation}\label{hmovingcoordslow}
u(t,x) \sim e^{i\phi_2^\pm} e^{i\nu_2 t} e^{i\kappa_2 \xi_2^\pm} U_{\rm H}(\xi_2^\pm) = u_2^\pm(t,x) , 
\qquad \xi_2=\const,\quad \varepsilon=\xi_2-\xi_1\rightarrow \pm\infty .
\end{equation}

\subsection{Moving-coordinate expansion of the Sasa-Satsuma oscillatory $2$-soliton}
For the Sasa-Satsuma oscillatory $2$-soliton solution from Proposition~\ref{thm:2soliton}, 
we apply the asymptotic expansions \eqref{fastexpansion} and \eqref{slowexpansion} 
to the functions \eqref{ssoscilX1}, \eqref{ssoscilX2}, \eqref{ssoscilY}. 

First using the expansion \eqref{fastexpansion} 
and neglecting subdominant terms for $\varepsilon\rightarrow \pm\infty$, 
we find 
\begin{align}
& 
\begin{aligned} 
e^{i\kappa_1\xi_1} V_{1\rm SS}(\xi_1,\xi_1+\varepsilon) & 
\sim e^{2k_2|\varepsilon|} \bigg( 
k_1(k_1^2+\kappa_1^2)^{1/4} |\kappa_1/2|^{1/2} |\kappa_2| e^{i\kappa_1\xi_1} e^{\pm 2 k_2\xi_1} 
\times
\\&\qquad\qquad\quad
\Big( \sqrt{\Delta\Gamma} e^{i(\alpha_2+\gamma_2)} e^{k_1\xi_1} 
+\frac{1}{\sqrt{\Delta\Gamma}} e^{-k_1\xi_1} e^{i(\gamma_2-\upsilon_2)} \Big) 
\bigg) 
+O(e^{k_2|\varepsilon|})
\end{aligned}
\label{ssX1fast}
\\
& e^{i\kappa_2(\xi_1+\varepsilon)}V_{2\rm SS}(\xi_1,\xi_1+\varepsilon)  
\sim O(e^{k_2|\varepsilon|})
\label{ssX2fast}
\\
& 
\begin{aligned} 
W_{\rm SS}(\xi_1,\xi_1+\varepsilon) & 
\sim e^{2k_2|\varepsilon|}\bigg( 
|\kappa_2| e^{\pm 2 k_2\xi_1} \Big( 
\tfrac{1}{2}|\kappa_1| \Big( \Delta\Gamma e^{\pm 2k_1\xi_1} + \frac{1}{\Delta\Gamma} e^{\mp 2k_1\xi_1} \Big)
\\&\qquad\qquad\quad
+(k_1^2+\kappa_1^2)^{1/2} \Big)
\bigg) +O(e^{k_2|\varepsilon|})
\end{aligned} 
\label{ssYfast}
\end{align}
(with $k_2>0$ due to equation \eqref{k1k2prop}). 
The expansion \eqref{ssYfast} can be expressed as 
\begin{equation}\label{ssdenomfast}
e^{-2k_2|\varepsilon|} e^{\mp 2k_2\xi_1} W_{\rm SS}(\xi_1,\xi_1+\varepsilon)  
\sim  
|\kappa_2|\big( \cosh(2k_1\xi_1^\pm) +(k_1^2+\kappa_1^2)^{1/2} \big), 
\qquad \varepsilon\rightarrow \pm\infty
\end{equation}
where
\begin{equation}\label{ssshiftcoord1}
\xi_1^\pm = \xi_1 \mp x_1, 
\quad
x_1 = -\frac{\ln(\Delta\Gamma)}{2k_1} >0
\end{equation}
is a shifted moving coordinate. 
Then writing the expansion \eqref{ssX1fast} in terms of this coordinate,
and using the relations \eqref{2solitonalphas}--\eqref{2solitonupsilons},
we get 
\begin{equation}\label{ssnumerfast}
\begin{aligned}
&  e^{-2k_2|\varepsilon|} e^{\mp 2k_2\xi_1} e^{i\kappa_1\xi_1} V_{1\rm SS}(\xi_1,\xi_1+\varepsilon)  
\\&\qquad 
\sim 
k_1(k_1^2+\kappa_1^2)^{1/4} |2\kappa_1|^{1/2} |\kappa_2| 
e^{\pm i(\gamma_2+\frac{1}{2}\delta_2 +\kappa_1x_1)} e^{i\kappa_1\xi_1^\pm} \cosh(k_1\xi_1^\pm + \tfrac{1}{2}i\lambda_1),
\qquad \varepsilon\rightarrow \pm\infty .
\end{aligned}
\end{equation}
These functions \eqref{ssdenomfast} and \eqref{ssnumerfast}
resemble the denominator and numerator 
in the oscillatory $1$-soliton solution \eqref{f1soliton} and \eqref{ssoscilfunct}
for the Sasa-Satsuma equation. 
Specifically, we have 
\begin{equation}\label{ssf1fast}
\begin{aligned}
e^{i\kappa_1\xi_1} V_{1\rm SS}(\xi_1,\xi_1+\varepsilon)/W_{\rm SS}(\xi_1,\xi_1+\varepsilon)
= \tilde f_{1\rm SS}(\xi_1,\xi_1+\varepsilon) 
\sim & e^{\pm i(\gamma_2+\frac{1}{2}\delta_2 +\kappa_1 x_1)} e^{i\kappa_1\xi_1^\pm} U_{\rm SS}(\xi_1^\pm), 
\\&\qquad \varepsilon\rightarrow \pm\infty
\end{aligned}
\end{equation}
where $\tilde f_{1\rm SS}$ is the function \eqref{f1soliton}
and $U_{\rm SS}$ is the function \eqref{ssoscilfunct}. 
Finally, 
the phase factor $e^{\pm i(\gamma_2+\frac{1}{2}\delta_2 +\kappa_1 x_1)}$ 
can be combined with $e^{i\phi_1}$ 
to get a shifted phase angle
\begin{equation}\label{ssshiftphi1}
\phi_1^\pm= \phi_1 \pm\eta_1 , 
\quad
\eta_1 = \gamma_2 +\tfrac{1}{2}\delta_2 +\kappa_1 x_1 .
\end{equation}
In a similar way, the expansion \eqref{ssX2fast} gives
\begin{equation}
e^{-2k_2|\varepsilon|} e^{\mp 2k_2\xi_1} e^{i\kappa_2(\xi_1+\varepsilon)} V_{2\rm SS}(\xi_1,\xi_1+\varepsilon)  
\sim 0 , 
\qquad \varepsilon\rightarrow \pm\infty
\end{equation}
whence 
\begin{equation}\label{ssf2fast}
e^{i\kappa_2(\xi_1+\varepsilon)} V_{2\rm SS}(\xi_1,\xi_1+\varepsilon)/W_{\rm SS}(\xi_1,\xi_1+\varepsilon)
= \tilde f_{2\rm SS}(\xi_1,\xi_1+\varepsilon) 
\sim 0 , 
\qquad \varepsilon\rightarrow \pm\infty .
\end{equation}

Combining equations \eqref{ssf1fast} and \eqref{ssf2fast} with equation \eqref{f2soliton}, 
we see that the asymptotic expansion of the Sasa-Satsuma oscillatory $2$-soliton solution
with respect to its leftmost coordinate is given by 
\begin{equation}\label{ssmovingcoordfast}
u(t,x) \sim e^{i\phi_1^\pm} e^{i\nu_1 t} e^{i\kappa_1\xi_1^\pm} U_{\rm SS}(\xi_1^\pm) = u_1^\pm(t,x) , 
\qquad \xi_1=\const,\quad \varepsilon=\xi_2-\xi_1\rightarrow \pm\infty . 
\end{equation}

Next using the expansion \eqref{slowexpansion} 
and neglecting subdominant terms for $\varepsilon\rightarrow \pm\infty$, 
we find 
\begin{align}
& 
e^{i\kappa_2(\xi_2-\varepsilon)} V_{1\rm SS}(\xi_2-\varepsilon,\xi_2) 
\sim O(e^{k_1|\varepsilon|})
\label{ssX1slow}
\\
& 
\begin{aligned} 
e^{i\kappa_2\xi_2} V_{2\rm SS}(\xi_2-\varepsilon,\xi_2) & 
\sim e^{2k_1|\varepsilon|} \bigg( 
k_2(k_2^2+\kappa_2^2)^{1/4} |\kappa_2/2|^{1/2} |\kappa_1| e^{i\kappa_2\xi_2} e^{\pm 2 k_1\xi_2} 
\times
\\&\qquad\qquad\quad
\Big( \sqrt{\Delta\Gamma} e^{-i(\alpha_1+\gamma_1)} e^{k_2\xi_2} 
+\frac{1}{\sqrt{\Delta\Gamma}} e^{-k_2\xi_2} e^{i(\upsilon_1-\gamma_1)} \Big) 
\bigg) 
+O(e^{k_1|\varepsilon|})
\end{aligned}
\label{ssX2slow}
\\
& 
\begin{aligned} 
W_{\rm SS}(\xi_2-\varepsilon,\xi_2) & 
\sim e^{2k_1|\varepsilon|}\bigg( 
|\kappa_1| e^{\pm 2 k_1\xi_2} \Big( 
\tfrac{1}{2}|\kappa_2| \Big( \Delta\Gamma e^{\mp 2k_2\xi_2} + \frac{1}{\Delta\Gamma} e^{\pm 2k_2\xi_2} \Big)
\\&\qquad\qquad\quad
+(k_2^2+\kappa_2^2)^{1/2} \Big)
\bigg) +O(e^{k_1|\varepsilon|})
\end{aligned} 
\label{ssYslow}
\end{align}
(with $k_1>0$ due to equation \eqref{k1k2prop}). 
We express the expansion \eqref{ssYslow} as 
\begin{equation}
e^{-2k_1|\varepsilon|} e^{\mp 2k_1\xi_2} W_{\rm SS}(\xi_2-\varepsilon,\xi_2) 
\sim  
|\kappa_1|\big( \cosh(2k_2\xi_2^\pm) +(k_2^2+\kappa_2^2)^{1/2} \big), 
\qquad \varepsilon\rightarrow \pm\infty
\end{equation}
where
\begin{equation}\label{ssshiftcoord2}
\xi_2^\pm = \xi_2 \mp x_2, 
\quad
x_2 = \frac{\ln(\Delta\Gamma)}{2k_2} <0
\end{equation}
is a shifted moving coordinate. 
Then we write the expansion \eqref{ssX2slow} in terms of this coordinate,
and use the relations \eqref{2solitonalphas}--\eqref{2solitonupsilons},
giving
\begin{equation}
\begin{aligned}
&  e^{-2k_1|\varepsilon|} e^{\mp 2k_1\xi_2} e^{i\kappa_2\xi_2} V_{2\rm SS}(\xi_2-\varepsilon,\xi_2) 
\\&\qquad
\sim 
k_2(k_2^2+\kappa_2^2)^{1/4} |2\kappa_2|^{1/2} |\kappa_1| e^{i\kappa_2\xi_2}
e^{\mp i(\gamma_1+\frac{1}{2}\delta_1 -\kappa_2 x_2)} e^{i\kappa_2\xi_2^\pm} \cosh(k_2\xi_2^\pm + \tfrac{1}{2}i\lambda_2),
\qquad\varepsilon\rightarrow \pm\infty . 
\end{aligned}
\end{equation}
Hence we have 
\begin{equation}\label{ssf2slow}
\begin{aligned}
e^{i\kappa_2\xi_2} V_{2\rm SS}(\xi_2-\varepsilon,\xi_2) /W_{\rm SS}(\xi_2-\varepsilon,\xi_2) 
= \tilde f_{2\rm SS}(\xi_2-\varepsilon,\xi_2) 
\sim & e^{\mp i(\gamma_1+\frac{1}{2}\delta_1 -\kappa_2 x_2)} e^{i\kappa_2\xi_2^\pm} U_{\rm SS}(\xi_2^\pm) , 
\\& \qquad \varepsilon\rightarrow \pm\infty . 
\end{aligned}
\end{equation}
The phase factor $e^{\mp i(\gamma_1+\frac{1}{2}\delta_1 -\kappa_2 x_2)}$ 
can be combined with $e^{i\phi_2}$ 
to get a shifted phase angle
\begin{equation}\label{ssshiftphi2}
\phi_2^\pm= \phi_2 \pm\eta_2 , 
\quad
\eta_2 = -\gamma_1-\tfrac{1}{2}\delta_1 +\kappa_2 x_2 . 
\end{equation}
Similarly, the expansion \eqref{ssX1slow} gives
\begin{equation}
e^{-2k_1|\varepsilon|} e^{\mp 2k_1\xi_2} e^{i\kappa_2(\xi_2-\varepsilon)} V_{1\rm SS}(\xi_2-\varepsilon,\xi_2) 
\sim 0 , 
\qquad \varepsilon\rightarrow \pm\infty 
\end{equation}
whence 
\begin{equation}\label{ssf1slow}
e^{i\kappa_2(\xi_2-\varepsilon)} V_{1\rm SS}(\xi_2-\varepsilon,\xi_2)/W_{\rm SS}(\xi_2-\varepsilon,\xi_2) 
= \tilde f_{1\rm SS}(\xi_2-\varepsilon,\xi_2) 
\sim 0 , 
\qquad \varepsilon\rightarrow \pm\infty . 
\end{equation}

Combining equations \eqref{ssf2slow} and \eqref{ssf1slow} with equation \eqref{f2soliton}, 
we see that the asymptotic expansion of the Sasa-Satsuma oscillatory $2$-soliton solution
with respect to its rightmost coordinate is given by 
\begin{equation}\label{ssmovingcoordslow}
u(t,x) \sim e^{i\phi_2^\pm} e^{i\nu_2 t} e^{i\kappa_2 \xi_2^\pm} U_{\rm SS}(\xi_2^\pm) = u_2^\pm(t,x) , 
\qquad \xi_2=\const,\quad \varepsilon=\xi_2-\xi_1\rightarrow \pm\infty . 
\end{equation}

\subsection{Asymptotic expansion for large time}
The precise correspondence between 
the moving coordinate expansion given by $\varepsilon\rightarrow\pm\infty$ 
and an asymptotic expansion $t\rightarrow\pm\infty$
will now be explained. 
In particular, 
through equations \eqref{movingcoords} and \eqref{asymptparm}, 
we will determine how large $|t|$ must so that 
the expansions \eqref{hmovingcoordfast} and \eqref{hmovingcoordslow} 
derived for the Hirota oscillatory $2$-soliton
and the expansions \eqref{ssmovingcoordfast} and \eqref{ssmovingcoordslow} 
derived for the Sasa-Satsuma oscillatory $2$-soliton
are approximately valid over some interval in $x$
at a finite time $-\infty<t<\infty$. 

From equations \eqref{hX1fast}--\eqref{hYfast}
and equations \eqref{ssX1fast}--\eqref{ssYfast}, 
we see that the expansions \eqref{hmovingcoordfast} and \eqref{ssmovingcoordfast} 
remain approximately valid if $\pm k_2\xi_2\gg 1$ holds, 
with $k_1|\xi_1^\pm| =O(1)$. 
These two conditions can be expressed explicitly as conditions on $t,x$ 
after we use equations \eqref{movingcoords}, \eqref{asymptparm}, \eqref{hshiftcoord1} and \eqref{ssshiftcoord1}
to get $\xi_2= \xi_1^\pm \pm x_1 +(c_1-c_2)t$. 
Then the condition $\pm k_2\xi_2\gg 1$ gives
\begin{equation}\label{fastlargecond}
\pm(\xi_1^\pm+(c_1-c_2)t) \gg \frac{1}{k_2} - x_1 
\end{equation}
while the other condition $k_1|\xi_1^\pm| =O(1)$ implies 
\begin{equation}\label{fastrangecond}
\pm\xi_1^\pm \gtrsim -\frac{1}{k_1} . 
\end{equation}
We now combine these two inequalities \eqref{fastlargecond} and \eqref{fastrangecond}, 
yielding 
\begin{equation}\label{tfastcond}
\pm t \gg \frac{\dfrac{1}{k_2} + \dfrac{1}{k_1} - x_1}{c_1-c_2}
\end{equation}
which determines the minimum size of $t$. 
Finally, from inequality \eqref{fastrangecond}, we have
\begin{equation}\label{xfastrange}
c_1t \pm x_1-\frac{1}{k_1} \lesssim x \lesssim c_1t \pm x_1 + \frac{1}{k_1} 
\end{equation}
which determines the interval in which $x$ lies. 
These are the conditions on $t,x$ under which 
the expansions \eqref{hmovingcoordfast} and \eqref{ssmovingcoordfast} 
approximately hold, 
giving
\begin{equation}\label{u1asypt}
u(t,x) \simeq e^{i\phi_1^\pm} e^{i\nu_1 t} \tilde f(x-c_1t \mp x_1) = u_1^\pm(t,x) . 
\end{equation}

There are similar conditions for the expansions \eqref{hmovingcoordfast} and \eqref{ssmovingcoordfast} 
to hold approximately from equations \eqref{hX1slow}--\eqref{hYslow}
and equations \eqref{ssX1slow}--\eqref{ssYslow}, 
so thus 
\begin{equation}\label{u2asypt}
u(t,x) \simeq e^{i\phi_2^\pm} e^{i\nu_2 t} \tilde f(x-c_2t \mp x_2) = u_2^\pm(t,x) . 
\end{equation}
We see that these equations remain approximately valid if 
$\mp k_1\xi_1\gg 1$ holds, 
with $k_2|\xi_2^\pm| =O(1)$.  
By using equations \eqref{movingcoords} and \eqref{asymptparm}, 
we get 
$\xi_1= \xi_2^\pm \pm x_2 -(c_1-c_2)t$. 
The condition $\mp k_1\xi_1\gg 1$ thereby gives
\begin{equation}\label{slowlargecond}
\mp(\xi_2^\pm-(c_1-c_2)t) \gg \frac{1}{k_1} +x_2
\end{equation}
while the other condition $k_2|\xi_2^\pm| =O(1)$ implies 
\begin{equation}\label{slowrangecond}
\mp\xi_2^\pm \gtrsim -\frac{1}{k_2} . 
\end{equation}
Then, combining these two inequalities \eqref{slowlargecond} and \eqref{slowrangecond}, 
we obtain 
\begin{equation}\label{tslowcond}
\pm t \gg \frac{\dfrac{1}{k_2} + \dfrac{1}{k_1} + x_2}{c_1-c_2}
\end{equation}
which determines the minimum size of $t$. 
Then from inequality \eqref{slowrangecond}, we have
\begin{equation}\label{xslowrange}
c_2t \pm x_2-\frac{1}{k_2} \lesssim x \lesssim c_2t \pm x_2 + \frac{1}{k_2} 
\end{equation}
which determines the interval in which $x$ lies. 

An important observation now is that the approximate expansions \eqref{u1asypt} and \eqref{u2asypt}
will hold simultaneously if $t$ satisfies both conditions 
\eqref{tfastcond} and \eqref{tslowcond}. 
Since equations \eqref{hshiftcoord1}, \eqref{hshiftcoord2}, 
\eqref{ssshiftcoord1}, \eqref{ssshiftcoord2} 
show that $x_1>0$ and $x_2<0$ in all cases, 
a simple sufficient condition on $t$ is given by 
\begin{equation}\label{tlargecond}
|t| \gg \frac{k_1 + k_2}{k_1k_2(c_1-c_2)} . 
\end{equation}

Another useful observation is that the previous analysis holds 
independently of the signs of $c_1$ and $c_2$,
including cases when one of $c_1$ or $c_2$ is zero.
Hence, we have established the following results. 

\begin{lem}\label{prop:h2solitonlarget}
For $t$ satisfying the condition \eqref{tlargecond}, 
the Hirota oscillatory $2$-soliton solution 
\eqref{2soliton}, \eqref{hoscilX1}--\eqref{hoscilY} 
with parameters 
$\phi_1$, $\phi_2$, $\nu_1$, $\nu_2$, $c_1>c_2$ 
has the form of an asymptotic superposition $u\simeq u_1^\pm + u_2^\pm$ 
in which $u_1^\pm$ and $u_2^\pm$ are distinct oscillatory waves
having respective speeds $c_1$ and $c_2$, 
temporal frequencies $\nu_1$ and $\nu_2$, 
phase angles $\phi_1^\pm$ and $\phi_2^\pm$ 
given by expressions \eqref{hshiftphi1} and \eqref{hshiftphi2}, 
and having positions that are determined by the respective moving coordinates 
\eqref{hshiftcoord1} and \eqref{hshiftcoord2}. 
\end{lem}

\begin{lem}\label{prop:ss2solitonlarget}
For $t$ satisfying the condition \eqref{tlargecond}, 
the Sasa-Satsuma oscillatory $2$-soliton solution 
\eqref{2soliton}, \eqref{ssoscilX1}--\eqref{ssoscilY} 
with parameters 
$\phi_1$, $\phi_2$, $\nu_1$, $\nu_2$, $c_1>c_2$ 
has the form of an asymptotic superposition $u\simeq u_1^\pm + u_2^\pm$ 
in which $u_1^\pm$ and $u_2^\pm$ are distinct oscillatory waves
having respective speeds $c_1$ and $c_2$, 
temporal frequencies $\nu_1$ and $\nu_2$, 
phase angles $\phi_1^\pm$ and $\phi_2^\pm$ 
given by expressions \eqref{ssshiftphi1} and \eqref{ssshiftphi2}, 
and having positions that are determined by the respective moving coordinates 
\eqref{ssshiftcoord1} and \eqref{ssshiftcoord2}. 
\end{lem}

When these oscillatory $2$-soliton solutions 
for the Hirota equation and the Sasa-Satsuma equation
have either $c_1=0$ or $c_2=0$,  
then the respective asymptotic wave $u_1^\pm$ or $u_2^\pm$ as $t\rightarrow \pm\infty$
is a standing wave.

\section{Constants of Motion}
\label{com}

For the Hirota equation \eqref{hmkdveq}
and the Sasa-Satsuma equation \eqref{ssmkdveq}, 
we recall that the conserved integrals defining 
momentum, energy, and Galilean energy 
are given by \cite{conslaws} 
(up to arbitrary normalization factors) 
\begin{align}
& \mathcal{P}= \int_{-\infty}^{+\infty} |u|^2\; dx
\label{mom}\\
& \mathcal{E}= \int_{-\infty}^{+\infty} 3(|u_x|^2 -4|u|^4)\; dx
\label{ener}\\
& \mathcal{C}= \int_{-\infty}^{+\infty} 3t(|u_x|^2 -4|u|^4)-x|u|^2\; dx
\label{galener}
\end{align}
which yield constants of motion for 
all smooth solutions $u(t,x)$ with sufficiently rapid decay 
$u\rightarrow 0$ as $x\rightarrow\pm\infty$. 
These integrals are related to the center of momentum defined by 
\EQ
\mathcal{X}(t)= \frac{1}{\mathcal{P}} \int_{-\infty}^{+\infty} x |u|^2\; dx
= \mathcal{X}(0) +\frac{\mathcal{E}}{\mathcal{P}} t
\label{comom}
\endEQ
where 
\EQ
\mathcal{C}= t\mathcal{E} -\mathcal{P}\mathcal{X}(t) 
= \mathcal{C}(0)= -\mathcal{P}\mathcal{X}(0) . 
\label{comomrel}
\endEQ
This is the same relation that holds for the corresponding 
constants of motion of the mKdV equation \cite{AncNgaWil}. 

The Hirota equation admits an additional conserved integral given by 
the angular twist \cite{conslaws}
(up to an arbitrary normalization factor) 
\begin{equation}
\mathcal{W}= \int_{-\infty}^{+\infty} \Re(i u\bar u_x)\; dx 
= -i\int_{-\infty}^{+\infty} |u|^2\arg(u)_x\; dx 
\label{twist}
\end{equation}
holding for all smooth solutions $u(t,x)$ with sufficiently rapid decay 
$u\rightarrow 0$ as $x\rightarrow\pm\infty$. 
This integral is not conserved for the Sasa-Satsuma equation. 

It is straightforward to evaluate these constants of motion explicitly 
for the oscillatory $1$-soliton solutions from Proposition~\ref{thm:1soliton} 
for the Hirota equation and the Sasa-Satsuma equation. 
For notional convenience we will denote 
\begin{equation}\label{betapm}
\beta_{\pm} = \cubrt{\sqrt{(c/3)^3 +(\nu/2)^2} \pm \nu/2} . 
\end{equation}

\begin{thm}\label{thm:com1soliton}
The Hirota oscillatory $1$-soliton 
\eqref{1soliton}, \eqref{hoscilfunct}
has angular twist, momentum, energy, and Galilean energy given by 
\begin{align}
& \mathcal{W}= \tfrac{1}{2}\kappa k
= \sqrt{3}(\beta{-}^2 -\alpha_+^2)/8
\label{htwi}
\\
& \mathcal{P}= \tfrac{1}{2} k
= \sqrt{3}(\beta{-} + \alpha_+)/4
\label{hmom}
\\
& \mathcal{E}= \tfrac{1}{2} k(k^2-3\kappa^2) 
= \sqrt{3}(\beta{-} + \alpha_+)c/4
\label{hener}
\\
& \mathcal{C}= 0 
\label{hgalener}
\end{align}
The Sasa-Satsuma oscillatory $1$-soliton 
\eqref{1soliton}, \eqref{ssoscilfunct}
has momentum, energy, and Galilean energy given by 
\begin{align}
& \mathcal{P}= k
= \sqrt{3}(\beta_{-} + \alpha_+)/2
\label{ssmom}
\\
& \mathcal{E}= k(k^2-3\kappa^2)
= \sqrt{3}(\beta_{-} + \alpha_+)c/2
\label{ssener}
\\
& \mathcal{C}= 0 
\label{ssgalener}
\end{align}
In both cases, the center of momentum is 
$\mathcal{X}(t) = ct$ with $c = \mathcal{E}/\mathcal{P}$. 
\end{thm}

We note that the center of momentum of these oscillatory $1$-solitons 
can be shifted arbitrarily by means of a space translation $x\rightarrow x-x_0$ 
applied to the moving coordinate $\xi=x-ct$ 
in equation \eqref{1soliton}, 
which leads to 
\begin{equation}
\mathcal{X}(t)= x_0+ ct . 
\end{equation}
This changes the Galilean energy 
\begin{equation}
\mathcal{C}= -x_0\mathcal{P}
\end{equation}
while the momentum and energy are unchanged. 

From the previous expressions, 
we can evaluate the momentum, energy, and Galilean energy of 
the oscillatory $2$-soliton solutions 
from Proposition~\ref{thm:2soliton} 
for the Hirota equation and the Sasa-Satsuma equation. 
In particular, 
we know from Lemmas~\ref{prop:h2solitonlarget} and~\ref{prop:ss2solitonlarget}
that each solution $u\sim u_1^\pm+u_2^\pm$ 
is asymptotically a superposition of two waves 
$u_1^\pm$ and $u_2^\pm$ as $t\rightarrow\pm\infty$.
Hence the conserved integrals \eqref{mom}, \eqref{ener}, \eqref{galener}
are respectively given by a sum of 
the momenta $\mathcal{P}_1$, $\mathcal{P}_2$,
the energies $\mathcal{E}_1$, $\mathcal{E}_2$,
and the Galilean energies $\mathcal{C}_1$, $\mathcal{C}_2$ 
associated with each individual wave. 
This yields the following result, using the notation \eqref{betapms}. 

\begin{thm}\label{thm:com2soliton}
The Hirota oscillatory $2$-soliton 
\eqref{2soliton}--\eqref{f2soliton}, \eqref{hoscilX1}--\eqref{hoscilY}
has angular twist, momentum, energy, and Galilean energy given by 
\begin{align}
& \mathcal{W}= \mathcal{W}_1 +\mathcal{W}_2
= \tfrac{1}{2}\kappa_1 k_1 + \tfrac{1}{2}\kappa_2 k_2
= \sqrt{3}(\beta_{1-}^2 -\beta_{1+}^2 + \beta_{2-}^2 -\beta_{2+}^2)/8
\label{h2solitontwi}
\\
&\mathcal{P}=\mathcal{P}_1 +\mathcal{P}_2
= \tfrac{1}{2}k_1+\tfrac{1}{2}k_2
= \sqrt{3}(\beta_{1-} + \beta_{1+} + \beta_{2-} + \beta_{2+})/4
\label{h2solitonmom}
\\
&\mathcal{E}=\mathcal{E}_1 +\mathcal{E}_2 
= \tfrac{1}{2} k_1(k_1^2-3\kappa_1^2) + \tfrac{1}{2} k_2(k_2^2-3\kappa_2^2) 
= \sqrt{3}((\beta_{1-} + \beta_{1+})c_1 + (\beta_{2-} + \beta_{2+})c_2)/4
\label{h2solitonener}
\\
&\mathcal{C}=\mathcal{C}_1 +\mathcal{C}_2 
= \mp(\tfrac{1}{2} k_1 x_1 + \tfrac{1}{2} k_2 x_2)
=0
\label{h2solitongalener}
\end{align}
where $x_1$ and $x_2$ are given by equations \eqref{hshiftcoord1} and \eqref{hshiftcoord2}. 
The Sasa-Satsuma oscillatory $2$-soliton 
\eqref{2soliton}--\eqref{f2soliton}, \eqref{ssoscilX1}--\eqref{ssoscilY}
has momentum, energy, and Galilean energy given by 
\begin{align}
&\mathcal{P}=\mathcal{P}_1 +\mathcal{P}_2
= k_1+k_2
= \sqrt{3}(\beta_{1-} + \beta_{1+} + \beta_{2-} + \beta_{2+})/2
\label{ss2solitonmom}
\\
&\mathcal{E}=\mathcal{E}_1 +\mathcal{E}_2 
= k_1(k_1^2-3\kappa_1^2) + k_2(k_2^2-3\kappa_2^2) 
= \sqrt{3}((\beta_{1-} + \beta_{1+})c_1 + (\beta_{2-} + \beta_{2+})c_2)/2
\label{ss2solitonener}
\\
&\mathcal{C}=\mathcal{C}_1 +\mathcal{C}_2 
= \mp( k_1 x_1 + k_2 x_2)
=0
\label{ss2solitongalener}
\end{align}
where $x_1$ and $x_2$ are given by equations \eqref{ssshiftcoord1} and \eqref{ssshiftcoord2}. 
In both cases, 
\begin{equation}\label{2solitoncomom}
\mathcal{X}(t) = \frac{\mathcal{E}}{\mathcal{P}} t 
\end{equation}
is the center of momentum, 
which moves at constant speed
\begin{equation}\label{2solitonspeed}
c=\frac{\mathcal{E}}{\mathcal{P}}
= \frac{\mathcal{P}_1 c_1 + \mathcal{P}_2 c_2}{\mathcal{P}_1 + \mathcal{P}_2}
= \frac{(\beta_{1-} + \beta_{1+})c_1 + (\beta_{2-} + \beta_{2+})c_2}{\beta_{1-} + \beta_{1+} + \beta_{2-} + \beta_{2+}} . 
\end{equation}
\end{thm}

\section{Position shifts and phase shifts}
\label{results} 

As shown by Lemmas~\ref{prop:h2solitonlarget} and~\ref{prop:ss2solitonlarget},
in the asymptotic past $t\rightarrow -\infty$ and future $t\rightarrow \infty$, 
the Hirota and Sasa-Satsuma oscillatory $2$-soliton solutions 
given in Proposition~\ref{thm:2soliton} 
reduce to a superposition 
$u\sim u_1^\pm + u_2^\pm$ of oscillatory $1$-solitons $u_1^\pm$ and $u_2^\pm$
having speeds $c_1$, $c_2$, temporal frequencies $\nu_1$, $\nu_2$, 
phase angles $\phi_1^\pm$, $\phi_2^\pm$, 
and having centers of momentum 
$\chi_1^\pm(t)=c_1 t\pm x_1$, $\chi_2^\pm(t)=c_2 t\pm x_2$,
with $c_1\neq c_2$.
Without loss of generality, 
we will assume $c_1>c_2$ hereafter, 
since $u$ is symmetric under simultaneously interchanging 
$c_1 \longleftrightarrow c_2$, $\nu_1 \longleftrightarrow \nu_2$, 
$\phi_1 \longleftrightarrow \phi_2$. 

We begin by examining some properties of 
the asymptotic oscillatory $1$-solitons 
\begin{equation}\label{asympt1soliton}
\begin{aligned}
& u_1^\pm=\exp(i\phi_1^\pm)\exp(i\nu_1 t)\exp(i\kappa_1\xi_1^\pm)U_1(\xi_1^\pm)
\\
& u_2^\pm=\exp(i\phi_2^\pm)\exp(i\nu_2 t)\exp(i\kappa_2\xi_2^\pm)U_2(\xi_2^\pm)
\end{aligned}
\end{equation}
where 
\begin{equation}\label{shiftcoords}
\xi_1^\pm =x-c_1t\mp x_1,
\quad 
\xi_2^\pm =x-c_2t\mp x_2
\end{equation}
are shifted moving coordinates, 
and where both $U_1$ and $U_2$ are given by the envelope function
\eqref{hoscilfunct} in the Hirota case
and \eqref{ssoscilfunct} in the Sasa-Satsuma case. 
First, 
the functions $U_1$ and $U_2$ are symmetric 
around $\xi_1^\pm=0$ and $\xi_2^\pm=0$, 
coinciding with the positions of the centers of momentum 
\begin{equation}\label{asymp1solitonposition}
x=\chi_1^\pm(t) =c_1 t\pm x_1 , 
\quad
x=\chi_2^\pm(t) =c_2 t\pm x_2
\end{equation}
for the two asymptotic oscillatory waves. 
Second, 
at these positions, 
the phase of both functions $U_1$ and $U_2$ vanishes, 
\begin{equation}\label{Uphase}
\arg(U_1)|_{\xi_1^\pm=0} = \arg(U_2)|_{\xi_2^\pm=0} 
=0 .
\end{equation}
Third, 
away from the positions \eqref{asymp1solitonposition},
the amplitude of the two asymptotic oscillatory waves has exponential decay
\begin{equation}\label{asymp1solitonamplitude}
\begin{aligned}
& |u_1^\pm| = |U_1| \sim O(\exp(-k_1 |\xi_1^\pm|)) ,
\qquad 
|\xi_1^\pm|\gg 1/k_1
\\
& |u_2^\pm| = |U_2| \sim O(\exp(-k_2 |\xi_2^\pm|)) , 
\qquad
|\xi_2^\pm|\gg 1/k_2
\end{aligned}
\end{equation}
while their phase has linear behaviour
\begin{equation}\label{asymp1solitonphase}
\begin{aligned}
& \arg(u_1^\pm) = \phi_1^\pm +\nu_1 t +\kappa_1\xi_1^\pm +\arg(U_1) 
\sim \psi_1 +\phi_1^\pm +\nu_1 t +\kappa_1\xi_1^\pm , 
\qquad
|\xi_1^\pm|\gg 1/k_1
\\
& \arg(u_2^\pm) = \phi_2^\pm +\nu_2 t +\kappa_2\xi_2^\pm +\arg(U_2)
\sim \psi_2 +\phi_2^\pm +\nu_2 t +\kappa_2\xi_2^\pm , 
\qquad
|\xi_2^\pm|\gg 1/k_2
\end{aligned}
\end{equation}
where 
$\psi_1=\psi_2=0$ 
in the Hirota case, 
and 
$\psi_1=\sgn(\xi_1^\pm)\arg( \kappa_1(\kappa_1+ik_1) )$, 
$\psi_2=\sgn(\xi_2^\pm)\arg( \kappa_2(\kappa_2+ik_2) )$ 
in the Sasa-Satsuma case. 
For graphical and analytical purposes, 
it will be more useful to work with the envelope phase of the two waves \eqref{asympt1soliton}. 

We recall that the envelope phase of an oscillatory wave 
$u = \exp(i\phi)\exp(i\nu t)\exp(i\kappa\xi)U(\xi)$
expressed in terms of a moving coordinate $\xi=x-ct-\chi_0$, 
with phase angle $\phi$, temporal frequency $\nu$, speed $c$, 
and center of momentum $\chi(t)=ct +\chi_0$,
is defined by \cite{AncMiaWil} 
\begin{equation}
\varphi(u) =\arg(u) -\kappa x -(\nu -\kappa c)t
= \phi -\kappa \chi_0+\arg(U) .
\end{equation}
Note 
$\arg(u)= \phi +\nu t + \kappa(x-ct-\chi_0)+\arg(U) = \kappa x +(\nu -\kappa c)t
+\varphi(u)$ is the total phase of $u$,
so thus $\varphi(u)$ represents the contribution to the phase of $u$
after the linear contributions $\kappa x -(\nu -\kappa c)t$
from the harmonic modulation are removed,
which corresponds to writing 
\begin{equation}\label{oscilenvelphase}
u = \exp(i\varphi)\exp(i(\kappa x -(\nu -\kappa c)t))|U(\xi)| .
\end{equation}
Applied to the asymptotic oscillatory waves \eqref{asympt1soliton}, 
this yields the envelope phases
$\varphi(u_1^\pm) = \arg(U_1)+\phi_1^\pm \mp\kappa_1 x_1$, 
$\varphi(u_2^\pm) = \arg(U_2)+\phi_2^\pm \mp\kappa_2 x_2$. 
The phase property \eqref{Uphase} shows that
\begin{equation}\label{asymp1solitonenvelphase}
\varphi(u_1^\pm)|_{\xi_1^\pm=0} = \phi_1^\pm \mp\kappa_1 x_1 = \varphi_1^\pm , 
\qquad
\varphi(u_2^\pm)|_{\xi_2^\pm=0} = \phi_2^\pm \mp \kappa_2 x_2 = \varphi_2^\pm , 
\end{equation}
where $\varphi_1^\pm = \phi_1\pm\gamma_2$, $\varphi_2^\pm = \phi_2\mp\gamma_1$, 
in the Hirota case, 
and $\varphi_1^\pm = \phi_1\pm(\gamma_2+\tfrac{1}{2}\delta_2)$, 
$\varphi_2^\pm = \phi_2\mp(\gamma_1+\tfrac{1}{2}\delta_1)$
in the Sasa-Satsuma case. 
Away from the center of momentum positions \eqref{asymp1solitonposition}, 
the envelope phases approach constant values 
\begin{equation}
\begin{aligned}
& \varphi(u_1^\pm)\sim \varphi_1^\pm +\psi_1,
\qquad
|\xi_1^\pm|\gg 1/k_1
\\
& \varphi(u_2^\pm)\sim \varphi_2^\pm +\psi_2,
\qquad
|\xi_2^\pm|\gg 1/k_2 . 
\end{aligned} 
\end{equation}

As $t\rightarrow \pm\infty$, 
the asymptotic positions \eqref{asymp1solitonposition} 
of the two oscillatory waves \eqref{asympt1soliton} 
lie on straight lines in the $(t,x)$-plane, 
with the lines $x=\chi_1^+(t)$ and $x=\chi_2^+(t)$ 
being each shifted relative to the lines $x=\chi_1^-(t)$ and $x=\chi_2^-(t)$ by 
a constant value 
\begin{equation}\label{asymptpositionshift}
\Delta x_1 = \chi_1^+(t) - \chi_1^-(t) =2x_1 , 
\quad
\Delta x_2 = \chi_2^+(t) - \chi_2^-(t) =2x_2 . 
\end{equation}
Likewise, the asymptotic phase angles of the two oscillatory waves 
are each shifted by a constant value 
\begin{equation}\label{asymptphaseshift}
\Delta \phi_1 = \phi_1^+ - \phi_1^- , 
\quad
\Delta \phi_2 = \phi_2^+ - \phi_2^- . 
\end{equation}
Hence the envelope phases also undergo shifts 
\begin{equation}\label{asymptenvelphaseshift}
\begin{aligned}
& \Delta \varphi_1 = \varphi(u_1^+) - \varphi(u_1^-) = \Delta\phi_1-\kappa_1\Delta x_1 = \varphi_1^+ - \varphi_1^- 
\\
& \Delta \varphi_2 = \varphi(u_2^+) - \varphi(u_2^-) = \Delta\phi_2-\kappa_2\Delta x_2  = \varphi_2^+ - \varphi_2^- 
\end{aligned}
\end{equation}
which are determined entirely by the asymptotic shifts 
\eqref{asymptpositionshift} and \eqref{asymptphaseshift}. 

From Lemmas~\ref{prop:h2solitonlarget} and~\ref{prop:ss2solitonlarget},
we have the following expressions for the shifts 
\eqref{asymptpositionshift} and \eqref{asymptenvelphaseshift}. 

\begin{thm}\label{thm:hshift}
For $t\rightarrow\pm\infty$ 
in the Hirota oscillatory $2$-soliton 
\eqref{2soliton}, \eqref{hoscilX1}--\eqref{hoscilY}, 
with $c_1>c_2$, 
the asymptotic soliton with speed $c_1$ and temporal frequency $\nu_1$
undergoes a shift in position and envelope phase given by 
\begin{gather}
\Delta x_1 = 
\frac{1}{k_1}\ln\bigg( 
\frac{(k_1 +k_2)^2+(\kappa_1 -\kappa_2)^2}{(k_1-k_2)^2+(\kappa_1-\kappa_2)^2} 
\bigg)
>0
\label{hpositionshift1}
\\
\Delta \varphi_1 = 
- 2\arg\big( (k_1+k_2)(k_1-k_2)+(\kappa_1-\kappa_2)^2 +i2k_2(\kappa_1-\kappa_2) \big) 
\label{hphaseshift1}
\end{gather}
while the asymptotic soliton with speed $c_2$ and temporal frequency $\nu_2$
undergoes a shift in position and envelope phase given by 
\begin{gather}
\Delta x_2 = 
-\frac{1}{k_2}\ln\bigg( 
\frac{(k_1 +k_2)^2+(\kappa_1 -\kappa_2)^2}{(k_1-k_2)^2+(\kappa_1-\kappa_2)^2} 
\bigg)
<0
\label{hpositionshift2}
\\
\Delta \varphi_2 = 
-2\arg\big( (k_1+k_2)(k_1-k_2)-(\kappa_1-\kappa_2)^2 + i2k_1(\kappa_1-\kappa_2) \big) 
\label{hphaseshift2}
\end{gather}
where $k_1,k_2,\kappa_1,\kappa_2$ are given in terms of $c_1,c_2,\nu_1,\nu_2$
by equations \eqref{rels1}--\eqref{betapms}.
\end{thm}

\begin{thm}\label{thm:ssshift}
For $t\rightarrow\pm\infty$ 
in the Sasa-Satsuma oscillatory $2$-soliton  
\eqref{2soliton}, \eqref{ssoscilX1}--\eqref{ssoscilY}, 
with $c_1>c_2$, $\nu_1\neq0$ and $\nu_2\neq0$, 
the asymptotic soliton with speed $c_1$ and temporal frequency $\nu_1$
undergoes a shift in position and envelope phase given by 
\begin{gather}
\Delta x_1 = 
\frac{1}{k_1}\ln\bigg( 
\frac{(k_1 +k_2)^2+(\kappa_1 -\kappa_2)^2}{(k_1-k_2)^2+(\kappa_1-\kappa_2)^2} 
\sqrt{\frac{(k_1 +k_2)^2+(\kappa_1 +\kappa_2)^2}{(k_1-k_2)^2+(\kappa_1+\kappa_2)^2}} 
\bigg)
>0
\label{sspositionshift1}
\\
\begin{aligned}
\Delta \varphi_1 = & 
-2\arg\big( (k_1+k_2)(k_1-k_2)+(\kappa_1-\kappa_2)^2 +i2k_2(\kappa_1-\kappa_2) \big) 
\\&\qquad
+ \arg\big( (k_1+k_2)(k_1-k_2) +(\kappa_1+\kappa_2)^2 +i2k_2(\kappa_1+\kappa_2) \big)
\end{aligned}
\label{ssphaseshift1}
\end{gather}
while the asymptotic soliton with speed $c_2$ and temporal frequency $\nu_2$
undergoes a shift in position and envelope phase given by 
\begin{gather}
\Delta x_2 = 
-\frac{1}{k_2}\ln\bigg( 
\frac{(k_1 +k_2)^2+(\kappa_1 -\kappa_2)^2}{(k_1-k_2)^2+(\kappa_1-\kappa_2)^2} 
\sqrt{\frac{(k_1 +k_2)^2+(\kappa_1 +\kappa_2)^2}{(k_1-k_2)^2+(\kappa_1+\kappa_2)^2}} 
\bigg)
<0
\label{sspositionshift2}
\\
\begin{aligned}
\Delta \varphi_2 = & 
- 2\arg\big( (k_1+k_2)(k_1-k_2)-(\kappa_1-\kappa_2)^2 +i2k_1(\kappa_1-\kappa_2) \big) 
\\&\qquad
+\arg\big( (k_1+k_2)(k_1-k_2) -(\kappa_1+\kappa_2)^2 -i2k_1(\kappa_1+\kappa_2) \big)
\end{aligned}
\label{ssphaseshift2}
\end{gather}
where $k_1,k_2,\kappa_1,\kappa_2$ are given in terms of $c_1,c_2,\nu_1,\nu_2$
by equations \eqref{rels1}--\eqref{betapms}.
\end{thm}

For both the Hirota and Sasa-Satsuma oscillatory $2$-solitons, 
as $t\rightarrow \pm\infty$ 
the centers of momentum and the phase angles of 
the two asymptotic oscillatory waves \eqref{asympt1soliton} 
are given by
\begin{equation}\label{asymptcomom}
\chi_1^\pm(t) = c_1t \pm \tfrac{1}{2}\Delta x_1, 
\quad
\chi_2^\pm(t) = c_2t \pm \tfrac{1}{2}\Delta x_2
\end{equation}
and 
\begin{equation}\label{asymptphases}
\phi_1^\pm = \phi_1 \pm \tfrac{1}{2}\Delta \phi_1, 
\quad
\phi_2^\pm = \phi_2 \pm \tfrac{1}{2}\Delta \phi_2 . 
\end{equation}

\subsection{Oscillatory wave collisions}

The Hirota and Sasa-Satsuma oscillatory $2$-soliton solutions 
\eqref{2soliton}--\eqref{f2soliton} 
describe a collision between two asymptotic oscillatory waves 
with speeds $c_1>c_2$ (or $c_1<c_2$). 
The collision is a right-overtake if $c_1>c_2\geq 0$ (or $c_2>c_1\geq 0$), 
a left-overtake if $0\geq c_1>c_2$ (or $0\geq c_2>c_1$), 
and a head-on if $c_1>0\geq c_2$ (or $c_2>0\geq c_1$). 
As will be now illustrated, 
in all cases the net effect of the collision is only to shift 
the asymptotic position and asymptotic phase angle of each wave,
where these shifts are given in Theorems~\ref{thm:hshift} and~\ref{thm:ssshift}.

The positions shifts are seen graphically 
in the asymptotic amplitude of the $2$-soliton solution, 
since for large $|t|$ we have 
\begin{equation}
|u| \sim \begin{cases} 
|u_1^\pm| = |U_1(\xi_1^\pm)|, & x\simeq \chi_1^\pm(t)
\\
|u_2^\pm| = |U_2(\xi_2^\pm)|, & x\simeq \chi_2^\pm(t)
\end{cases} 
\end{equation}
from Lemmas~\ref{prop:h2solitonlarget} and~\ref{prop:ss2solitonlarget},
where $\xi_1^\pm$ and $\xi_2^\pm$ are the shifted moving coordinates \eqref{shiftcoords}
which determine the positions \eqref{asymp1solitonposition} 
of the two asymptotic oscillatory waves, 
and where $U_1$ and $U_2$ are the envelope functions for these waves, 
given in Proposition~\ref{thm:1soliton}. 
Similarly, we have 
\begin{equation}\label{asymptphas}
\arg(u) \sim \begin{cases} 
\arg(u_1^\pm) = \varphi(u_1^\pm) +\kappa_1 x + (\nu_1 -c_1\kappa_1)t, 
& x\simeq \chi_1^\pm(t)
\\
\arg(u_2^\pm) = \varphi(u_2^\pm) +\kappa_2 x + (\nu_2 -c_1\kappa_2)t, 
& x\simeq \chi_2^\pm(t)
\end{cases} 
\end{equation}
yielding the asymptotic phase of the $2$-soliton solution. 

To see the phase shifts graphically, 
it is useful to remove the asymptotic linear part of $\arg(u)$ 
by defining an envelope phase for the $2$-soliton solution
similarly to the definition \eqref{oscilenvelphase}
for oscillatory waves \cite{AncMiaWil}. 
Consider, for a $2$-soliton solution, the factorization
\begin{equation}
u = \exp(i\varphi)\Big( 
\exp(i(\nu_1 t +\kappa_1\xi_1))|V_1(\xi_1,\xi_2)| +\exp(i(\nu_2 t +\kappa_2\xi_2))|V_2(\xi_1,\xi_2)| 
\Big) \frac{A(\xi_1,\xi_2,t)}{W(\xi_1,\xi_2)}
\end{equation}
where $V_1$, $V_2$, $W$ are the envelope functions 
in Proposition~\ref{thm:2soliton},
and where $A$ is an amplitude normalization factor. 
Equating this form for $u$ to the oscillatory form \eqref{2soliton}--\eqref{f2soliton}, 
we obtain the envelope phase 
\begin{equation}\label{envelphase}
\varphi(u) = \arctan\left( 
\frac{|V_1|\Im(\exp(i\phi_1)(V_1+\exp(-i\Phi)V_2)) + |V_2|\Im(\exp(i\phi_2)(V_2+\exp(i\Phi)V_1))}{|V_1|\Re(\exp(i\phi_1)(V_1+\exp(-i\Phi)V_2)) + |V_2|\Re(\exp(i\phi_2)(V_2+\exp(i\Phi)V_1))}
\right)
\end{equation}
with
\begin{equation}
\Phi= \phi_1+\nu_1 t +\kappa_1\xi_1 -\phi_2 - \nu_2 t -\kappa_2\xi_2
= \phi_1-\phi_2+ (\kappa_1-\kappa_2)x +(\nu_1-\nu_2 +c_2\kappa_2 -c_1\kappa_1)t . 
\end{equation}
The envelope phase \eqref{envelphase} 
essentially represents the contribution to the phase of $u$
after the linear contributions from the harmonic modulation of 
the two asymptotic oscillatory waves are removed. 
In particular, let 
\begin{equation}\label{thetas}
\theta_1 = |V_1|/(|V_1|+|V_2|) ,
\quad
\theta_2 = |V_2|/(|V_1|+|V_2|) 
\end{equation}
denote normalized envelope functions satisfying the properties 
\begin{gather}
\theta_1 + \theta_2 =1
\label{thetarel1}\\
\theta_1 \sim \begin{cases} 
1 , 
& x\simeq \chi_1^\pm(t)
\\
0 , 
& x\simeq \chi_2^\pm(t)
\end{cases} 
\label{thetarel2}\\
\theta_2 \sim \begin{cases} 
1 , 
& x\simeq \chi_2^\pm(t)
\\
0 , 
& x\simeq \chi_1^\pm(t)
\end{cases} 
\label{thetarel3}
\end{gather}
Then we can write the envelope phase as 
\begin{equation}
\varphi(u) = 
\arctan\left( 
\frac{\theta_1^2\sin(\sigma_1) +\theta_2^2\sin(\sigma_2) +\theta_1\theta_2(\sin(\sigma_2-\Phi)+ \sin(\sigma_1+\Phi))}{\theta_1^2\cos(\sigma_1) +\theta_2^2\cos(\sigma_2) +\theta_1\theta_2(\cos(\sigma_2-\Phi)+ \cos(\sigma_1+\Phi))}{}
\right)
\end{equation}
with 
\begin{equation}
\sigma_1 = \arg(V_1)+\phi_1,
\quad
\sigma_2 = \arg(V_2)+\phi_2 . 
\end{equation}
The properties \eqref{thetarel1}--\eqref{thetarel3} 
combined with the asymptotic phase \eqref{asymptphas}
show that 
\begin{equation}
\varphi(u) \sim \begin{cases} 
\varphi(u_1^\pm) , 
& x\simeq \chi_1^\pm(t)
\\
\varphi(u_2^\pm) , 
& x\simeq \chi_2^\pm(t)
\end{cases} 
\end{equation}
whereby the envelope phase of $u$ 
asymptotically matches the envelope phase \eqref{asymp1solitonenvelphase} 
of each asymptotic oscillatory wave. 

\begin{figure}[H]
\centering
\begin{subfigure}[t]{.5\textwidth}
\includegraphics[width=\textwidth]{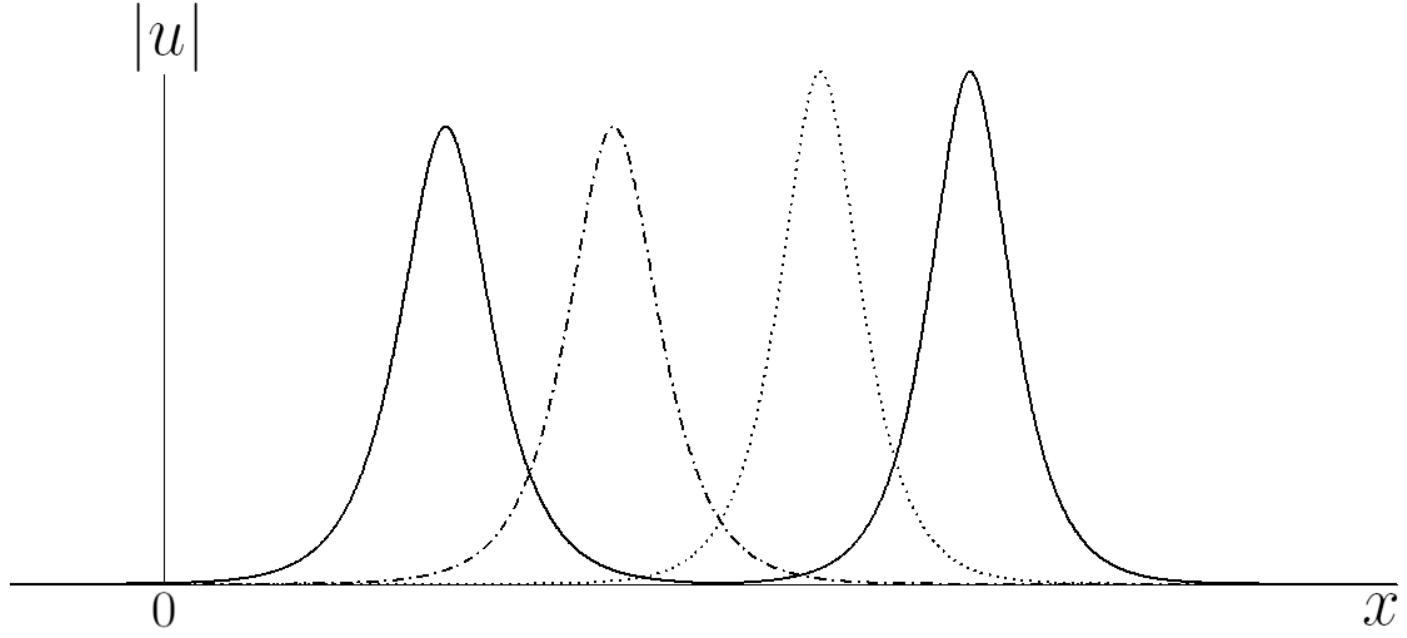}
\captionof{figure}{amplitude in asymptotic future}
\end{subfigure}%
\begin{subfigure}[t]{.5\textwidth}
\includegraphics[width=\textwidth]{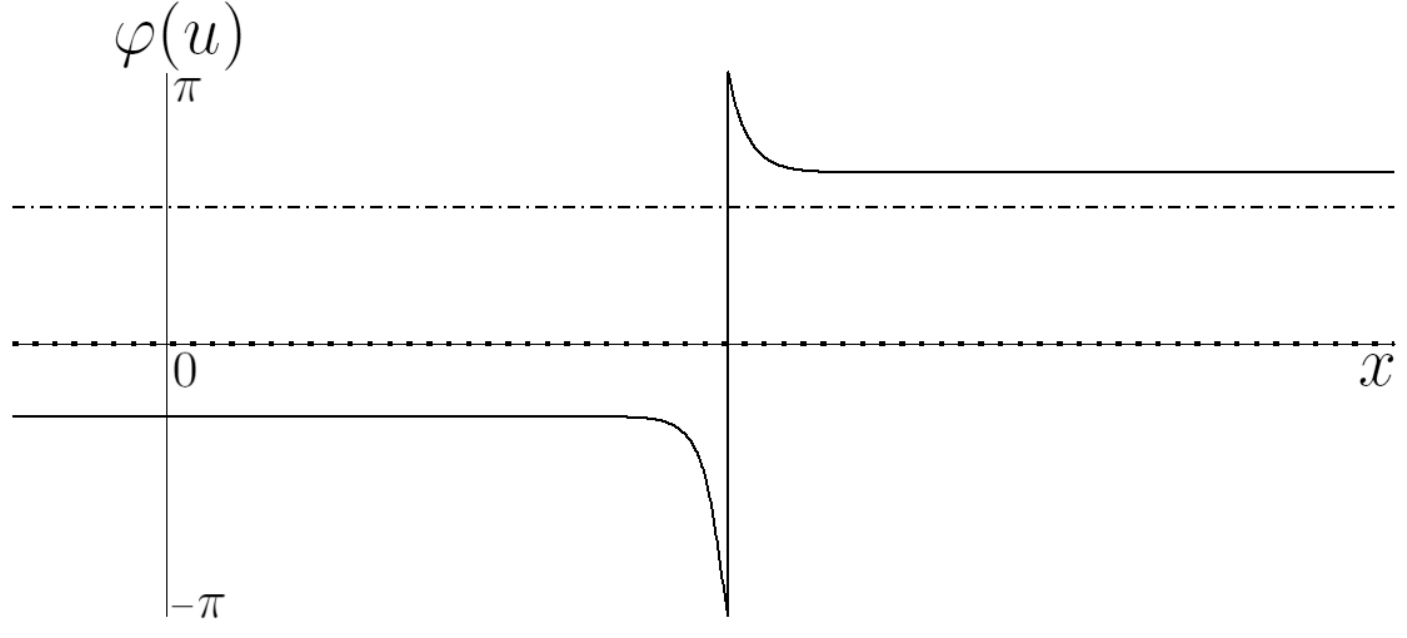}
\captionof{figure}{envelope phase in asymptotic future}
\end{subfigure}
\caption{Hirota oscillatory $2$-soliton right-overtake 
(in solid) and oscillatory $1$-solitons (in dots and dot-dash) 
with $c_1=4$, $c_2=2$, $\nu_1=2$, $\nu_2=5$, $\varphi_1^-=0$, $\varphi_2^-=\pi/2$}
\label{oscil_hirota_right-overtake-shifts_c1=4_c2=2_nu1=2_nu2=5_phi1=0_phi2=halfpi}
\end{figure}

\begin{figure}[H]
\centering
\begin{subfigure}[t]{.5\textwidth}
\includegraphics[width=\textwidth]{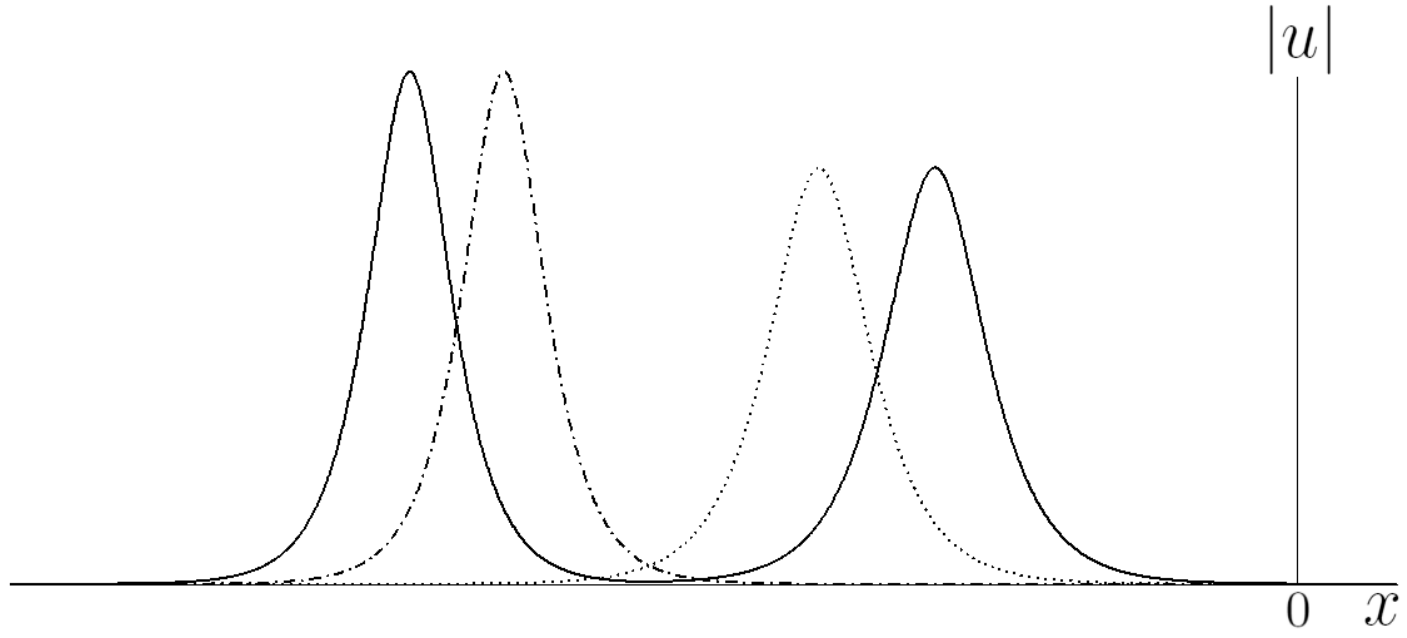}
\captionof{figure}{amplitude in asymptotic future}
\end{subfigure}%
\begin{subfigure}[t]{.5\textwidth}
\includegraphics[width=\textwidth]{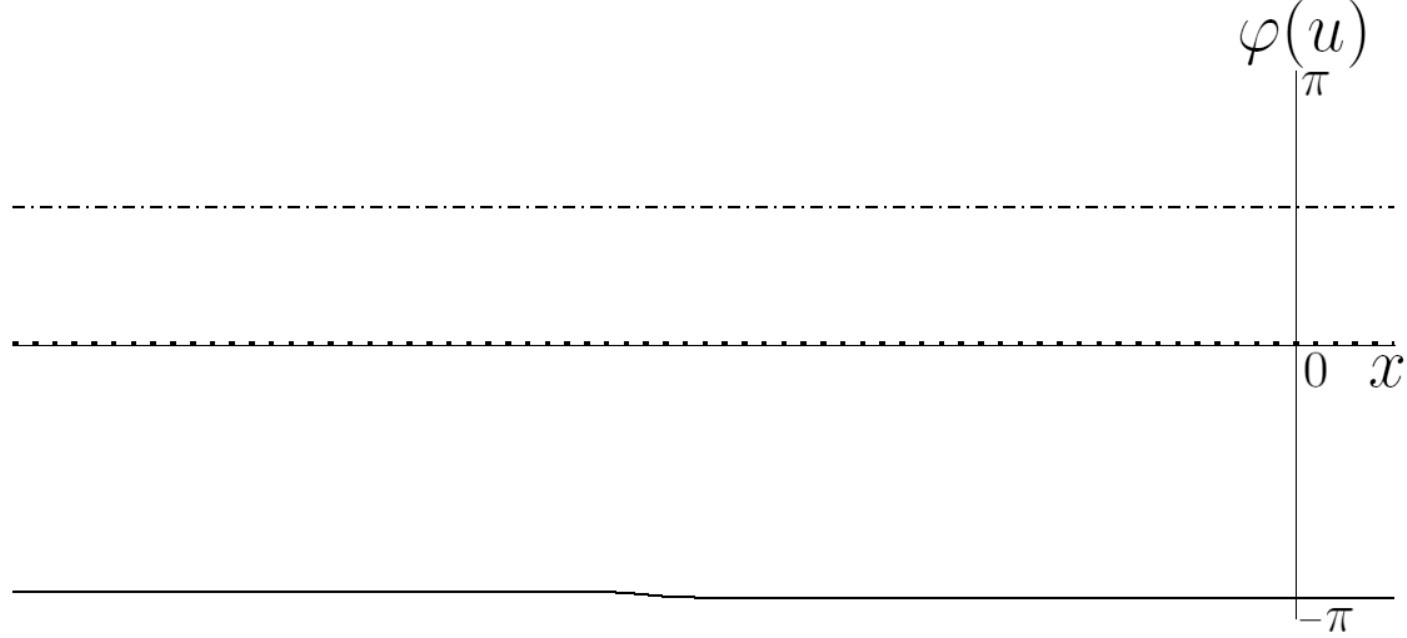}
\captionof{figure}{envelope phase in asymptotic future}
\end{subfigure}
\caption{Hirota oscillatory $2$-soliton left-overtake 
(in solid) and oscillatory $1$-solitons (in dots and dot-dash) 
with $c_1=-2$, $c_2=-4$, $\nu_1=2$, $\nu_2=5$, $\varphi_1^-=0$, $\varphi_2^-=\pi/2$}
\label{oscil_hirota_left-overtake-shifts_c1=-2_c2=-4_nu1=2_nu2=5_phi1=0_phi2=halfpi}
\end{figure}

\begin{figure}[H]
\centering
\begin{subfigure}[t]{.5\textwidth}
\includegraphics[width=\textwidth]{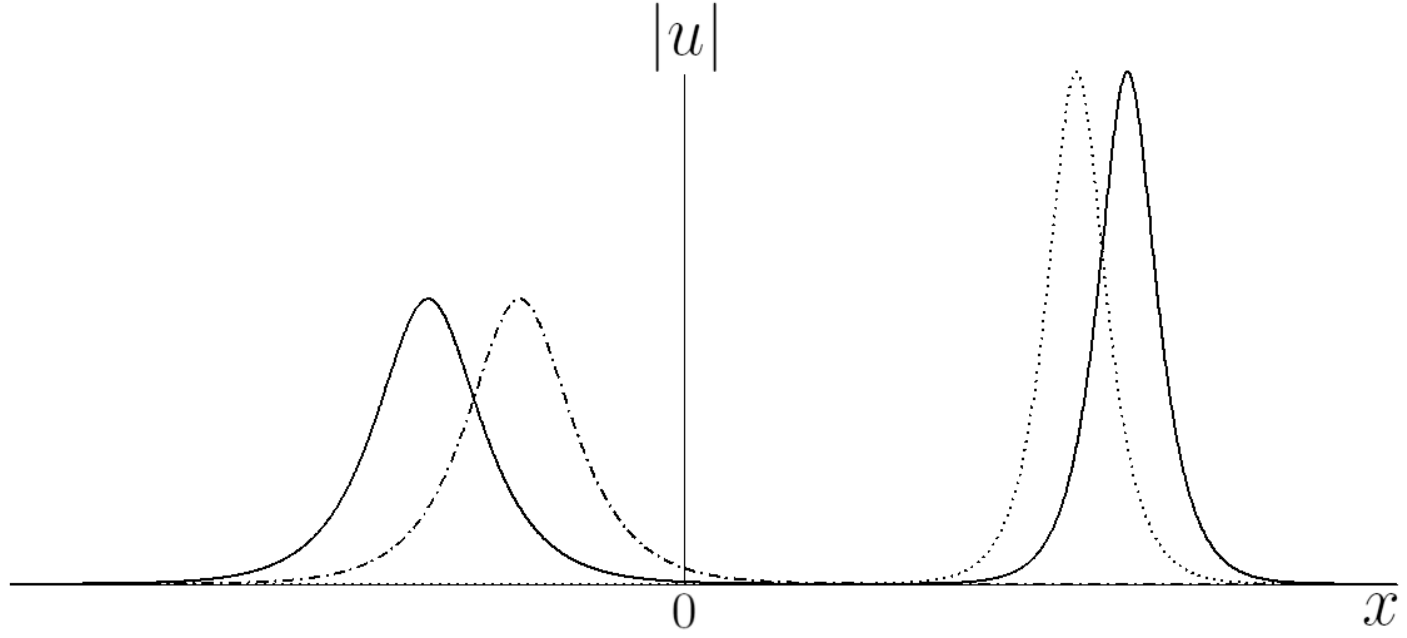}
\captionof{figure}{amplitude in asymptotic future}
\end{subfigure}%
\begin{subfigure}[t]{.5\textwidth}
\includegraphics[width=\textwidth]{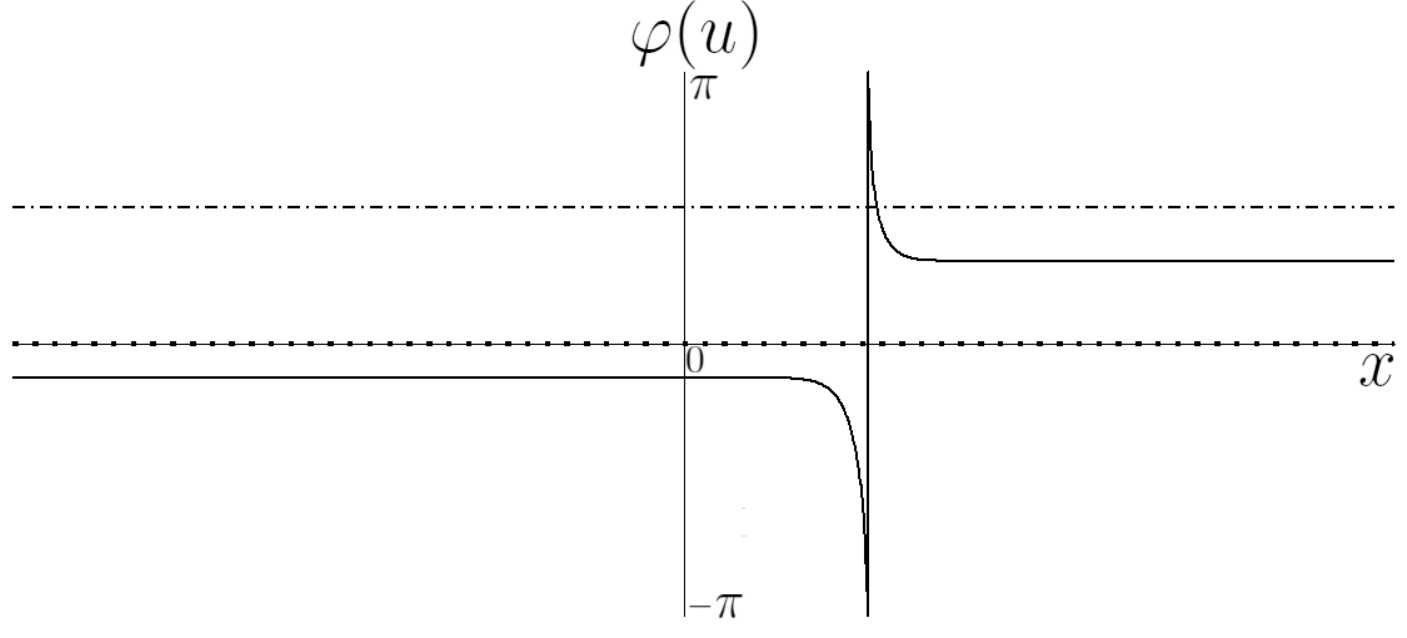}
\captionof{figure}{envelope phase in asymptotic future}
\end{subfigure}
\caption{Hirota oscillatory $2$-soliton head-on 
(in solid) and oscillatory $1$-solitons (in dots and dot-dash) 
with $c_1=4$, $c_2=-2$, $\nu_1=2$, $\nu_2=5$, $\varphi_1^-=0$, $\varphi_2^-=\pi/2$}
\label{oscil_hirota_headon-shifts_c1=4_c2=-2_nu1=2_nu2=5_phi1=0_phi2=halfpi}
\end{figure}

\begin{figure}[H]
\centering
\begin{subfigure}[t]{.5\textwidth}
\includegraphics[width=\textwidth]{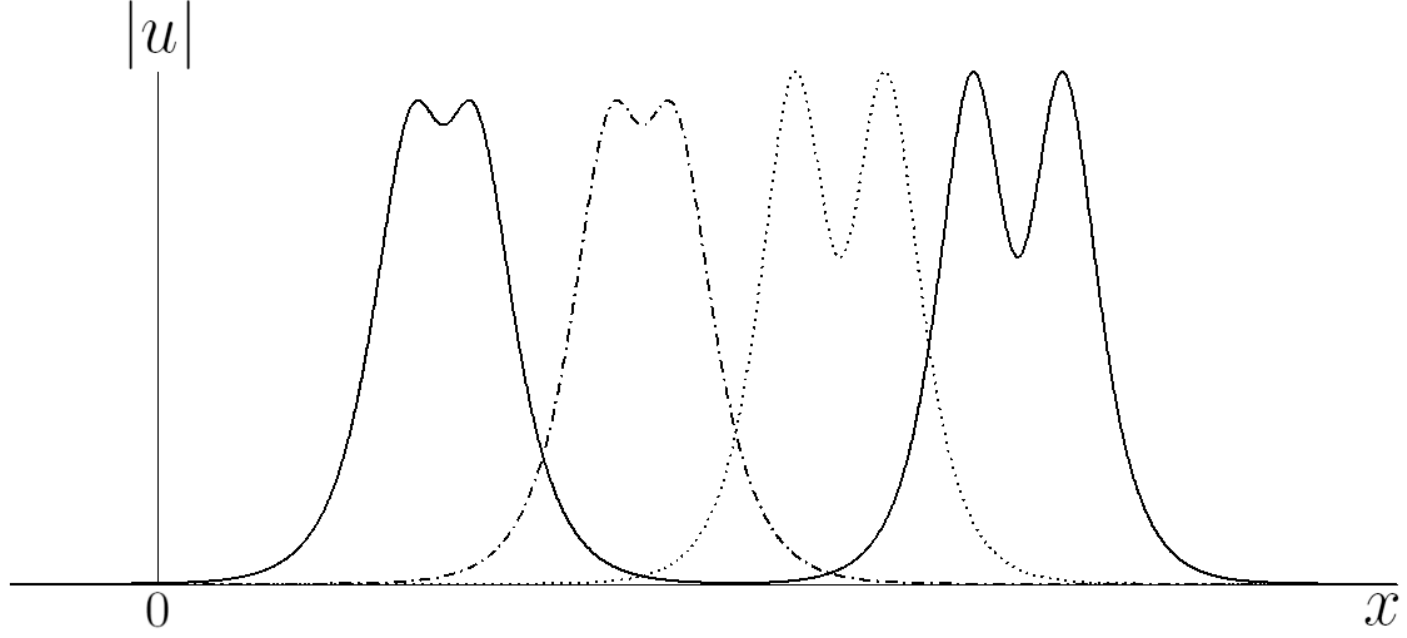}
\captionof{figure}{amplitude in asymptotic future}
\end{subfigure}%
\begin{subfigure}[t]{.5\textwidth}
\includegraphics[width=\textwidth]{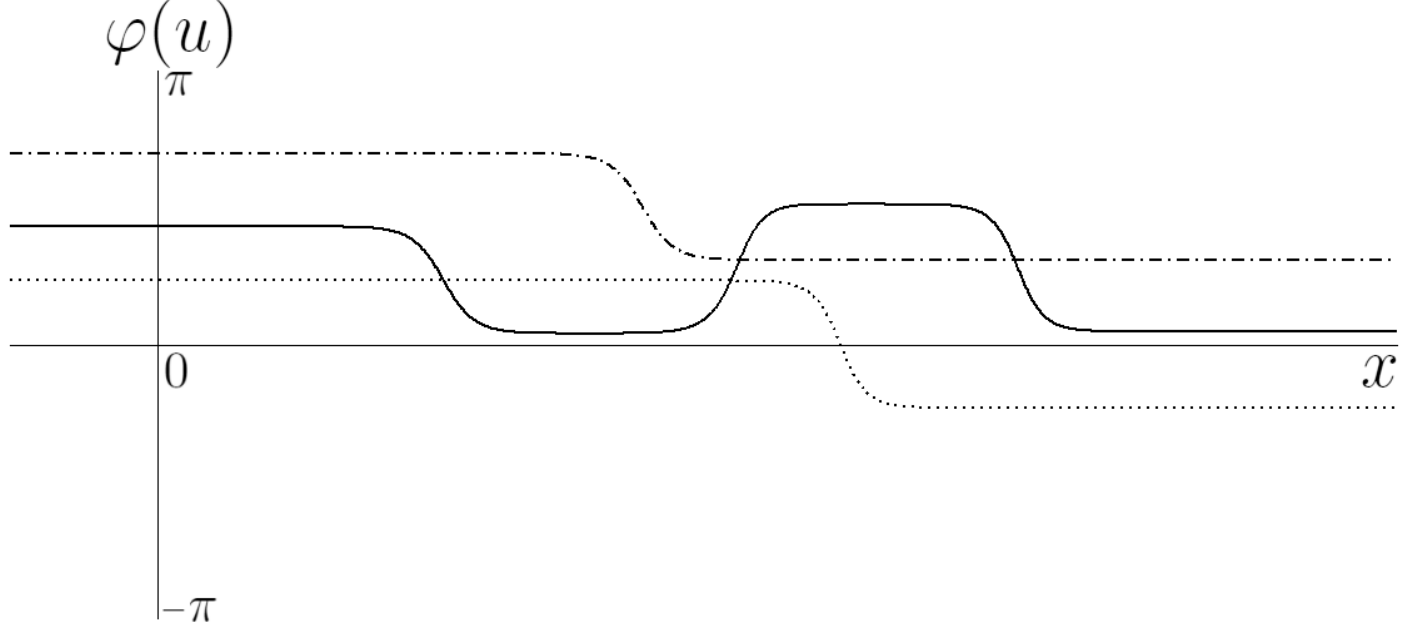}
\captionof{figure}{envelope phase in asymptotic future}
\end{subfigure}
\caption{Sasa-Satsuma oscillatory $2$-soliton right-overtake 
(in solid) and oscillatory $1$-solitons (in dots and dot-dash) 
with $c_1=4$, $c_2=2$, $\nu_1=2$, $\nu_2=5$, $\varphi_1^-=0$, $\varphi_2^-=\pi/2$}
\label{oscil_ss_right-overtake-shifts_c1=4_c2=2_nu1=2_nu2=5_phi1=0_phi2=halfpi}
\end{figure}

\begin{figure}[H]
\centering
\begin{subfigure}[t]{.5\textwidth}
\includegraphics[width=\textwidth]{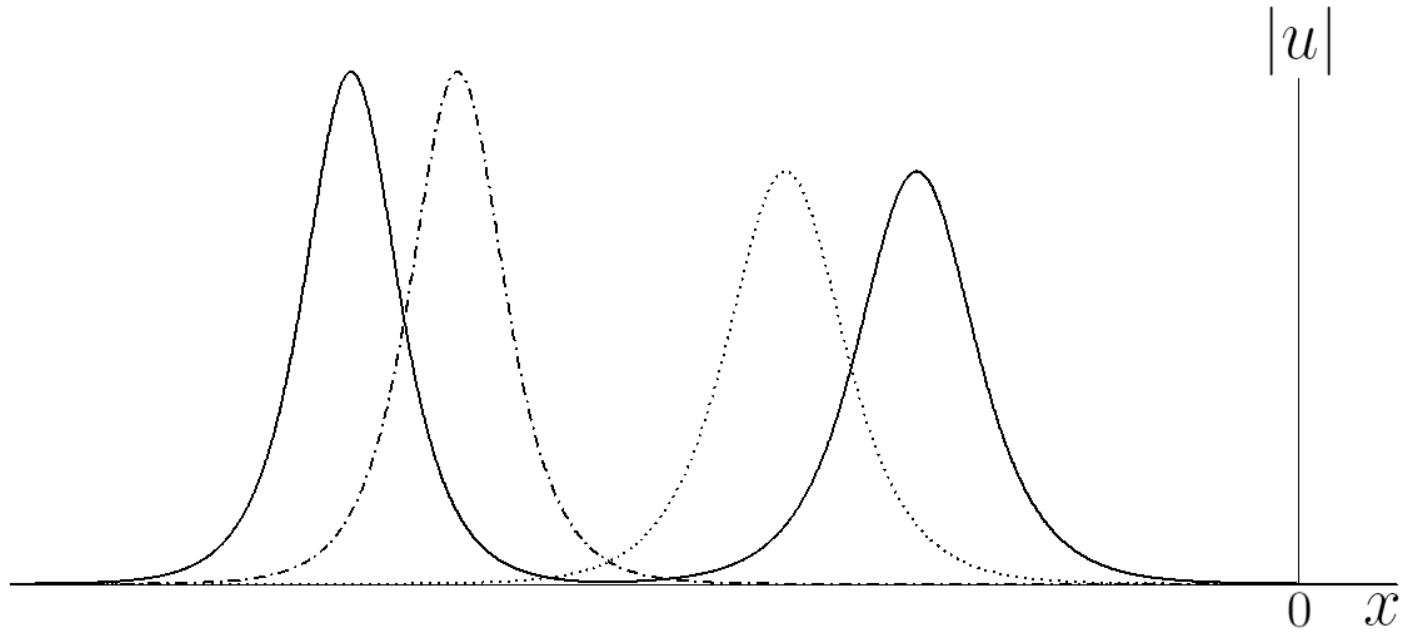}
\captionof{figure}{amplitude in asymptotic future}
\end{subfigure}%
\begin{subfigure}[t]{.5\textwidth}
\includegraphics[width=\textwidth]{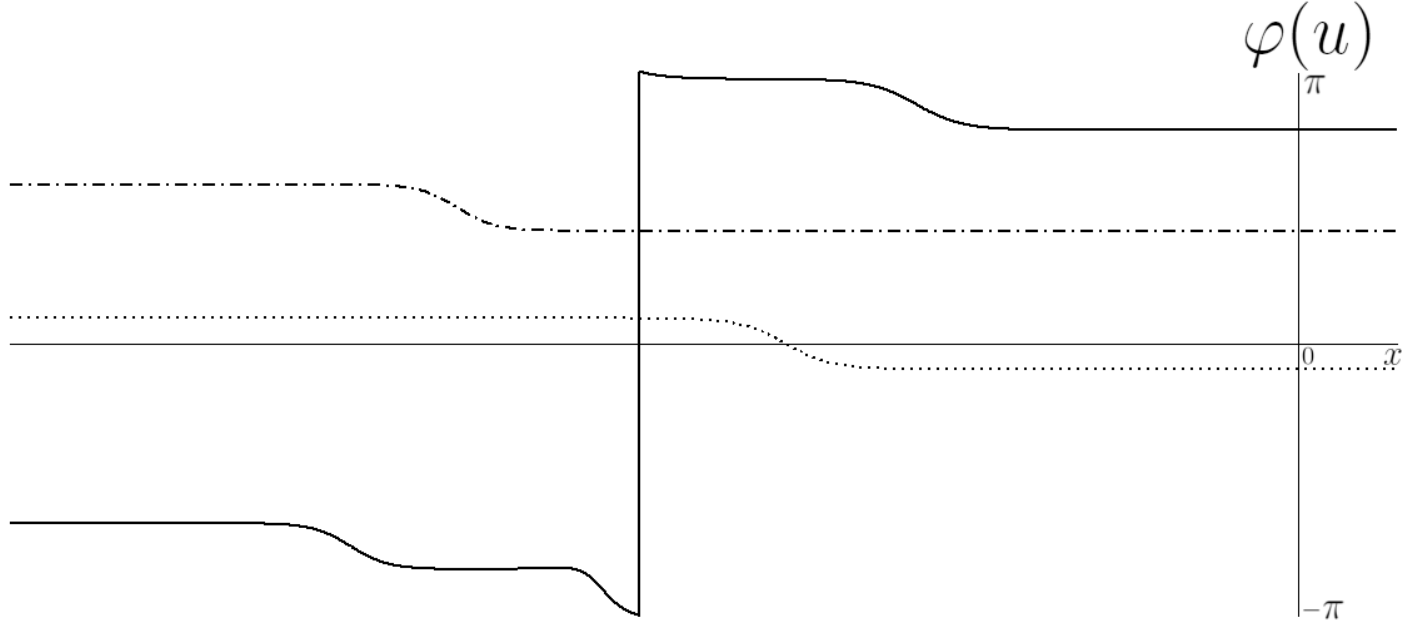}
\captionof{figure}{envelope phase in asymptotic future}
\end{subfigure}
\caption{Sasa-Satsuma oscillatory $2$-soliton left-overtake 
(in solid) and oscillatory $1$-solitons (in dots and dot-dash) 
with $c_1=-2$, $c_2=-4$, $\nu_1=2$, $\nu_2=5$, $\varphi_1^-=0$, $\varphi_2^-=\pi/2$}
\label{oscil_ss_left-overtake-shifts_c1=-2_c2=-4_nu1=2_nu2=5_phi1=0_phi2=halfpi}
\end{figure}

\begin{figure}[H]
\centering
\begin{subfigure}[t]{.5\textwidth}
\includegraphics[width=\textwidth]{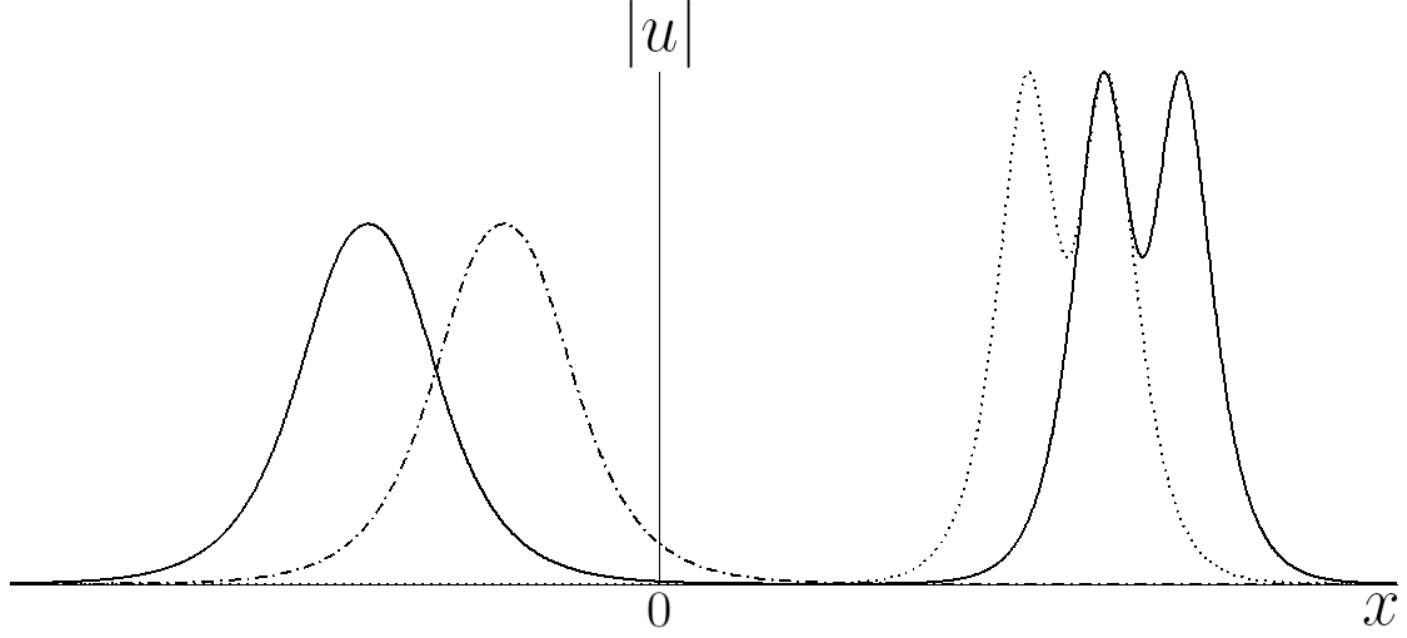}
\captionof{figure}{amplitude in asymptotic future}
\end{subfigure}%
\begin{subfigure}[t]{.5\textwidth}
\includegraphics[width=\textwidth]{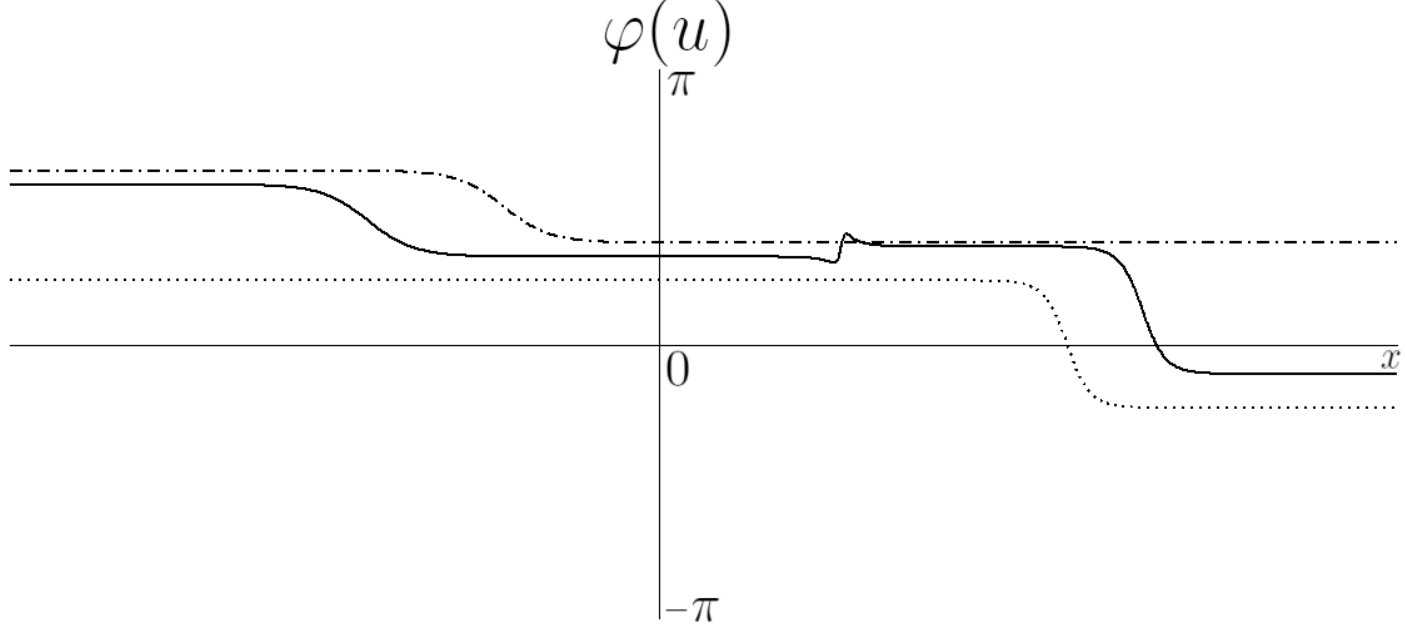}
\captionof{figure}{envelope phase in asymptotic future}
\end{subfigure}
\caption{Sasa-Satsuma oscillatory $2$-soliton head-on 
(in solid) and oscillatory $1$-solitons (in dots and dot-dash) 
with $c_1=4$, $c_2=-2$, $\nu_1=2$, $\nu_2=5$, $\varphi_1^-=0$, $\varphi_2^-=\pi/2$}
\label{oscil_ss_headon-shifts_c1=4_c2=-2_nu1=2_nu2=5_phi1=0_phi2=halfpi}
\end{figure}

The amplitude and envelope phase of 
the oscillatory $2$-soliton solutions \eqref{2soliton}--\eqref{f2soliton} 
are illustrated in 
\figref{oscil_hirota_right-overtake-shifts_c1=4_c2=2_nu1=2_nu2=5_phi1=0_phi2=halfpi}--
\figref{oscil_hirota_headon-shifts_c1=4_c2=-2_nu1=2_nu2=5_phi1=0_phi2=halfpi}
for the Hirota case,
and in 
\figref{oscil_ss_right-overtake-shifts_c1=4_c2=2_nu1=2_nu2=5_phi1=0_phi2=halfpi}--
\figref{oscil_ss_headon-shifts_c1=4_c2=-2_nu1=2_nu2=5_phi1=0_phi2=halfpi}
for the Sasa-Satsuma case. 

\subsection{Position shifts in oscillatory wave collisions}

In a right-overtake collision with $c_1>c_2>0$,
the asymptotic solitons $u_1^\pm$ and $u_2^\pm$ 
are oscillatory waves that each move to the right,
where $u_1^\pm$ is the faster wave and $u_2^\pm$ is the slower wave. 
The effect of the collision on the asymptotic positions of these waves 
is to shift the fast wave forward (\ie/ to the right, since $\Delta x_1>0$)
and the slow wave backward (\ie/ to the left, since $\Delta x_2<0$). 
Similarly, in a left-overtake collision with $0>c_1>c_2$,
the asymptotic solitons $u_1^\pm$ and $u_2^\pm$ 
are oscillatory waves that each move to the left,
where now $u_1^\pm$ is the slower wave and $u_2^\pm$ is the faster wave. 
The collision affects the asymptotic positions of the two waves 
by shifting the fast wave forward (\ie/ to the left, since $\Delta x_2<0$)
and the slow wave backward (\ie/ to the right, since $\Delta x_1>0$). 
In contrast, in a head-on collision with $c_1>0>c_2$,
the asymptotic soliton $u_1^\pm$ is a right-moving oscillatory wave 
while the other asymptotic soliton $u_2^\pm$ is a left-moving oscillatory wave. 
The collision has the effect that the asymptotic positions of both waves 
are shifted forward relative to their directions of motion, 
since the right-moving wave undergoes a shift to the right (due to $\Delta x_1>0$)
and the left-moving wave undergoes a shift to the left (due to $\Delta x_2<0$). 

In all cases, the asymptotic positions shifts $\Delta x_1$ and $\Delta x_2$
satisfy the algebraic relation
\begin{equation}\label{shiftrel}
k_1\Delta x_1 + k_2\Delta x_2 =0 
\end{equation}
which holds as a direct consequence of 
the oscillatory $2$-soliton solution 
having a center of momentum that moves at a constant speed, 
as shown by equation \eqref{2solitoncomom}.

\subsection{Standing waves}
An oscillatory $1$-soliton \eqref{oscil1soliton} with $c=0$ and $\nu\neq0$
is a time-periodic standing wave. 
The standing wave solutions for the Hirota equation \eqref{hmkdveq}
and the Sasa-Satsuma equation \eqref{ssmkdveq} 
are presented in oscillatory form in Proposition~\ref{thm:1soliton}. 

Collisions of an oscillatory wave with a standing wave 
are described by the oscillatory $2$-soliton \eqref{2soliton} 
when $c_1=0$, $c_2\neq 0$, $\nu_1\neq 0$,
or when $c_2=0$, $c_1\neq 0$, $\nu_2\neq 0$. 
These collision solutions for the Hirota and Sasa-Satsuma equations 
are special cases of the solutions presented in Proposition~\ref{thm:2soliton}. 
They have not previously appeared in the literature. 

\begin{figure}[H]
\centering
\begin{subfigure}[t]{.5\textwidth}
\includegraphics[width=\textwidth]{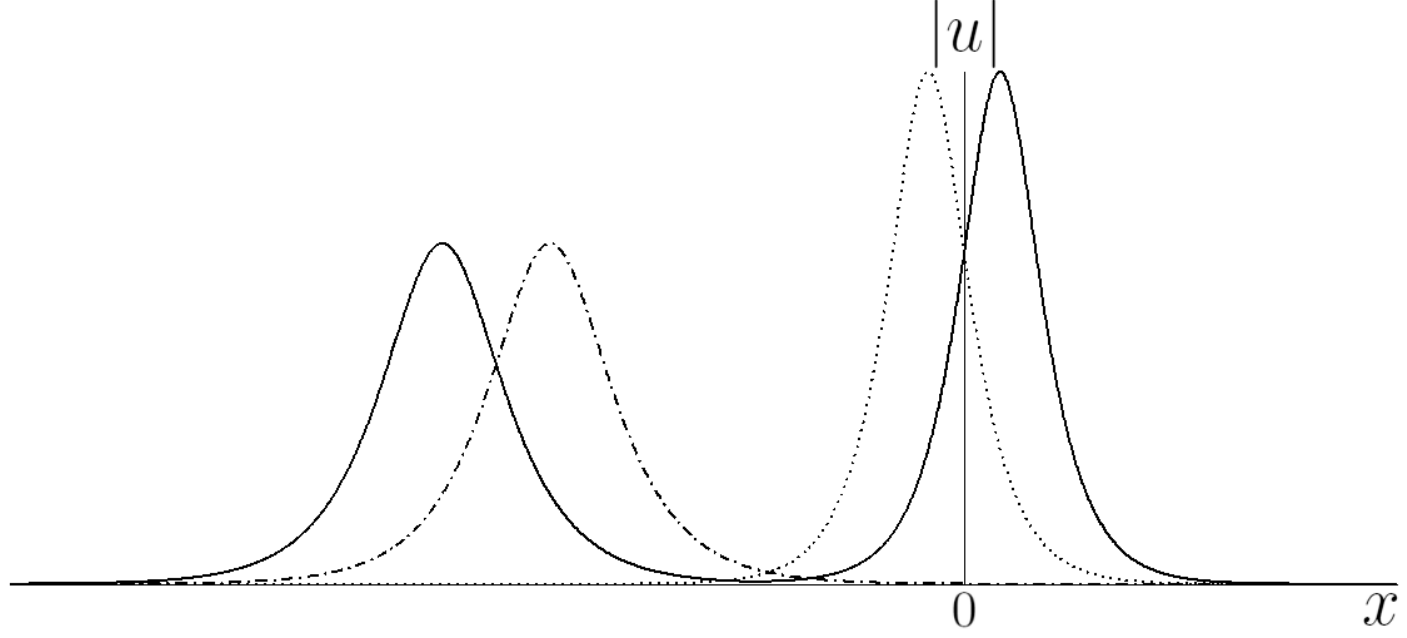}
\captionof{figure}{amplitude in asymptotic future}
\end{subfigure}%
\begin{subfigure}[t]{.5\textwidth}
\includegraphics[width=\textwidth]{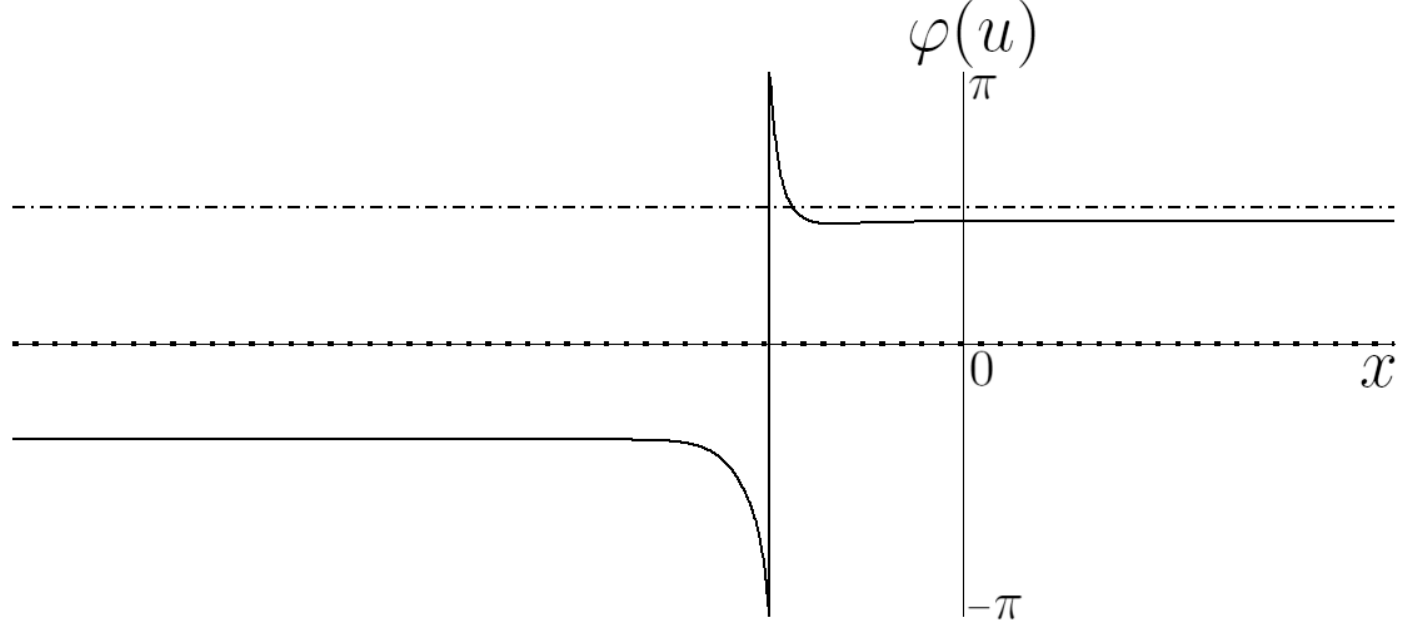}
\captionof{figure}{envelope phase in asymptotic future}
\end{subfigure}
\caption{Hirota $2$-soliton collision (in solid) of 
a left-moving oscillatory wave (in dot-dash) and a standing wave (in dots)
with $c_1=0$, $c_2=-4$, $\nu_1=2$, $\nu_2=5$, $\varphi_1^-=0$, $\varphi_2^-=\pi/2$}
\label{standing_hirota_left-overtake_shifts_c1=0_c2=-4_nu1=2_nu2=5_phi1=0_phi2=halfpi}
\end{figure}

\begin{figure}[H]
\centering
\begin{subfigure}[t]{.5\textwidth}
\includegraphics[width=\textwidth]{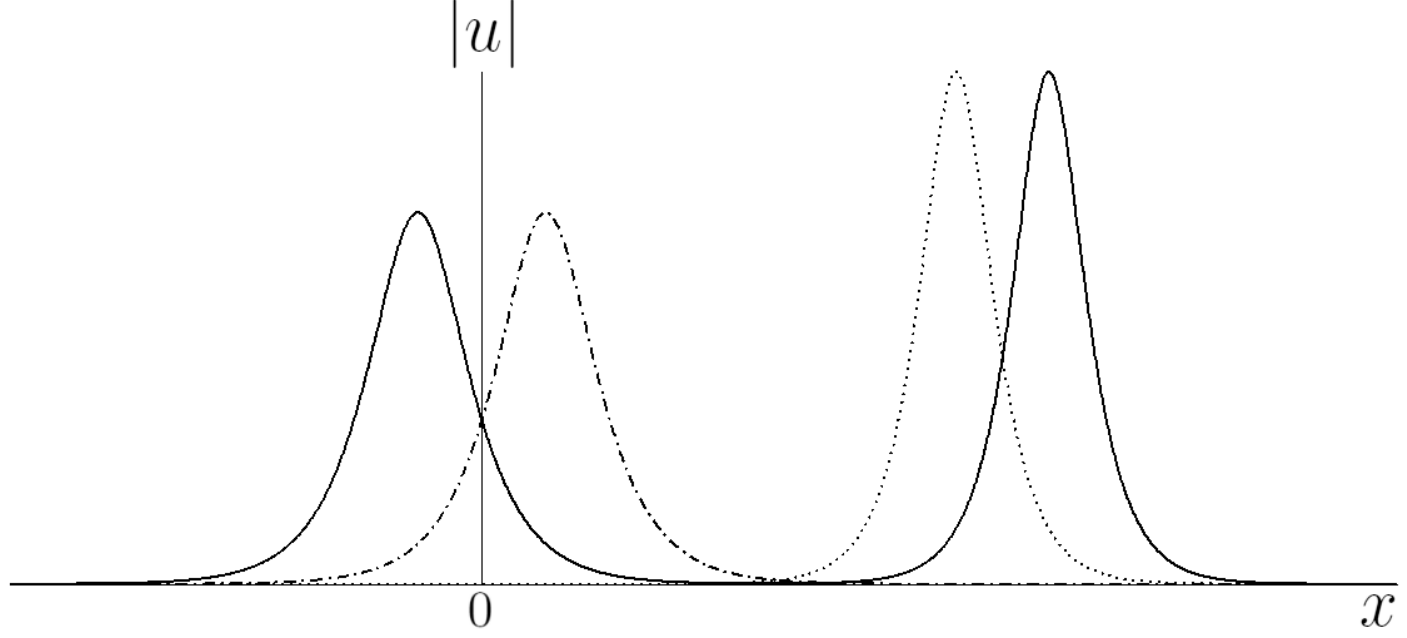}
\captionof{figure}{amplitude in asymptotic future}
\end{subfigure}%
\begin{subfigure}[t]{.5\textwidth}
\includegraphics[width=\textwidth]{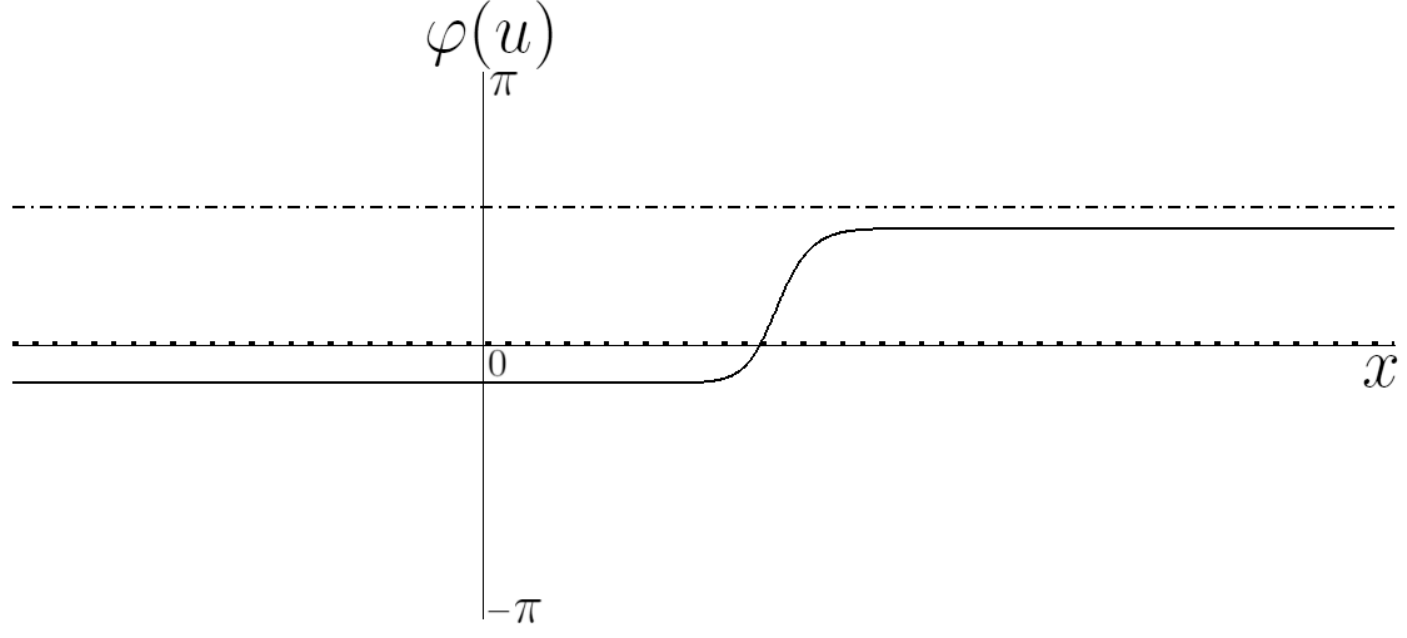}
\captionof{figure}{envelope phase in asymptotic future}
\end{subfigure}
\caption{Hirota $2$-soliton collision (in solid) of 
a right-moving oscillatory soliton (in dots) and a standing wave (in dot-dash)
with $c_1=4$, $c_2=0$, $\nu_1=2$, $\nu_2=5$, $\varphi_1^-=0$, $\varphi_2^-=\pi/2$}
\label{standing_hirota_right-overtake_shifts_c1=4_c2=0_nu1=2_nu2=5_phi1=0_phi2=halfpi}
\end{figure}

\begin{figure}[H]
\centering
\begin{subfigure}[t]{.5\textwidth}
\includegraphics[width=\textwidth]{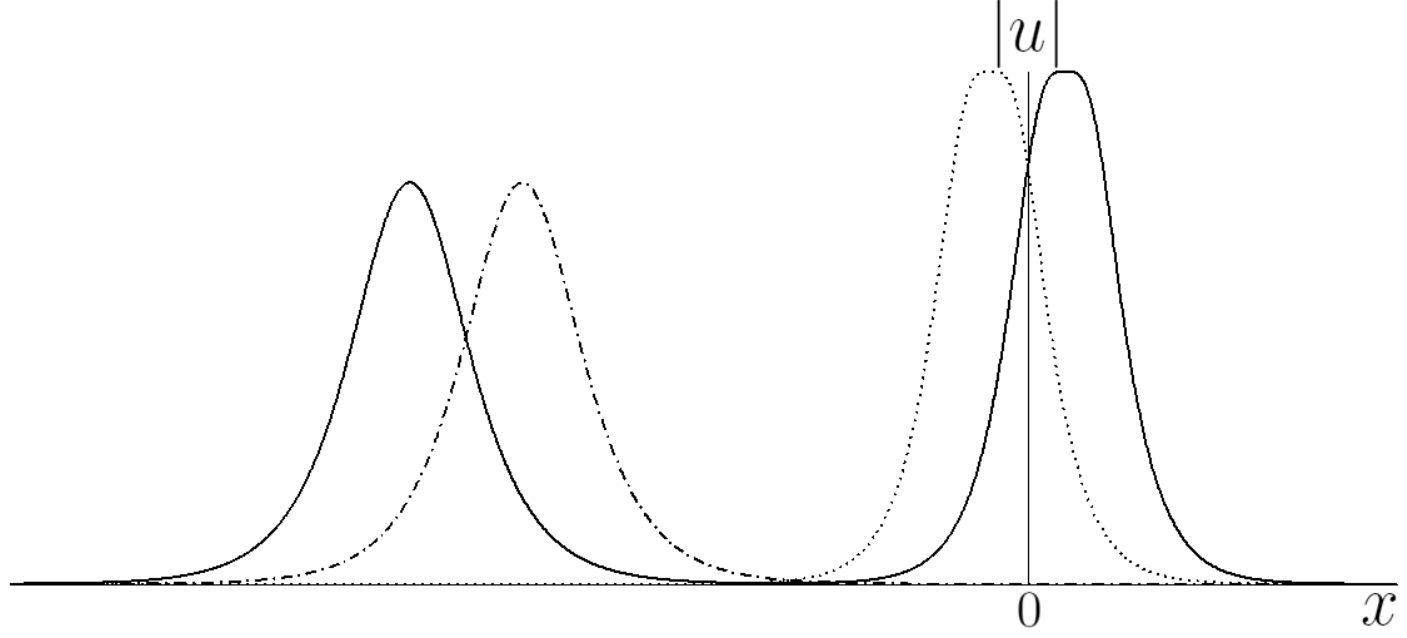}
\captionof{figure}{amplitude in asymptotic future}
\end{subfigure}%
\begin{subfigure}[t]{.5\textwidth}
\includegraphics[width=\textwidth]{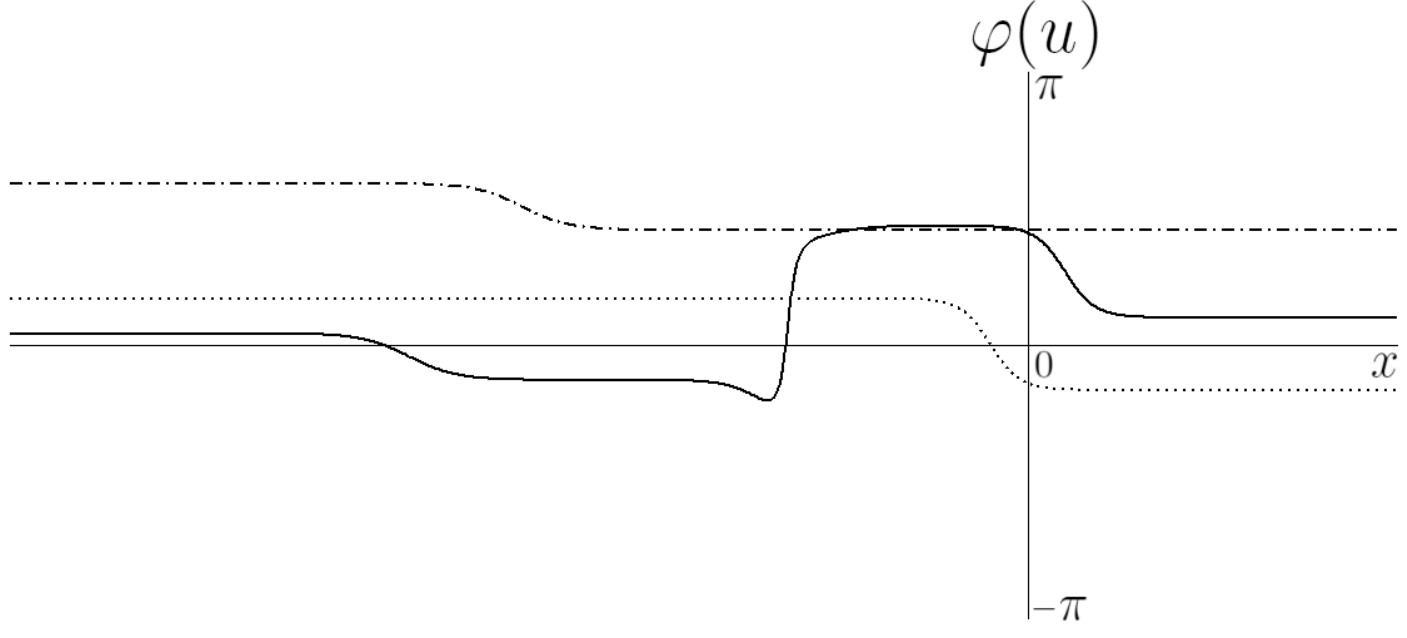}
\captionof{figure}{envelope phase in asymptotic future}
\end{subfigure}
\caption{Sasa-Satsuma collision (in solid) of 
a left-moving oscillatory soliton (in dot-dash) and a standing-wave soliton (in dots)
with $c_1=0$, $c_2=-4$, $\nu_1=2$, $\nu_2=5$, $\varphi_1^-=0$, $\varphi_2^-=\pi/2$}
\label{standing_ss_left-overtake_shifts_c1=0_c2=-4_nu1=2_nu2=5_phi1=0_phi2=halfpi}
\end{figure}

\begin{figure}[H]
\centering
\begin{subfigure}[t]{.5\textwidth}
\includegraphics[width=\textwidth]{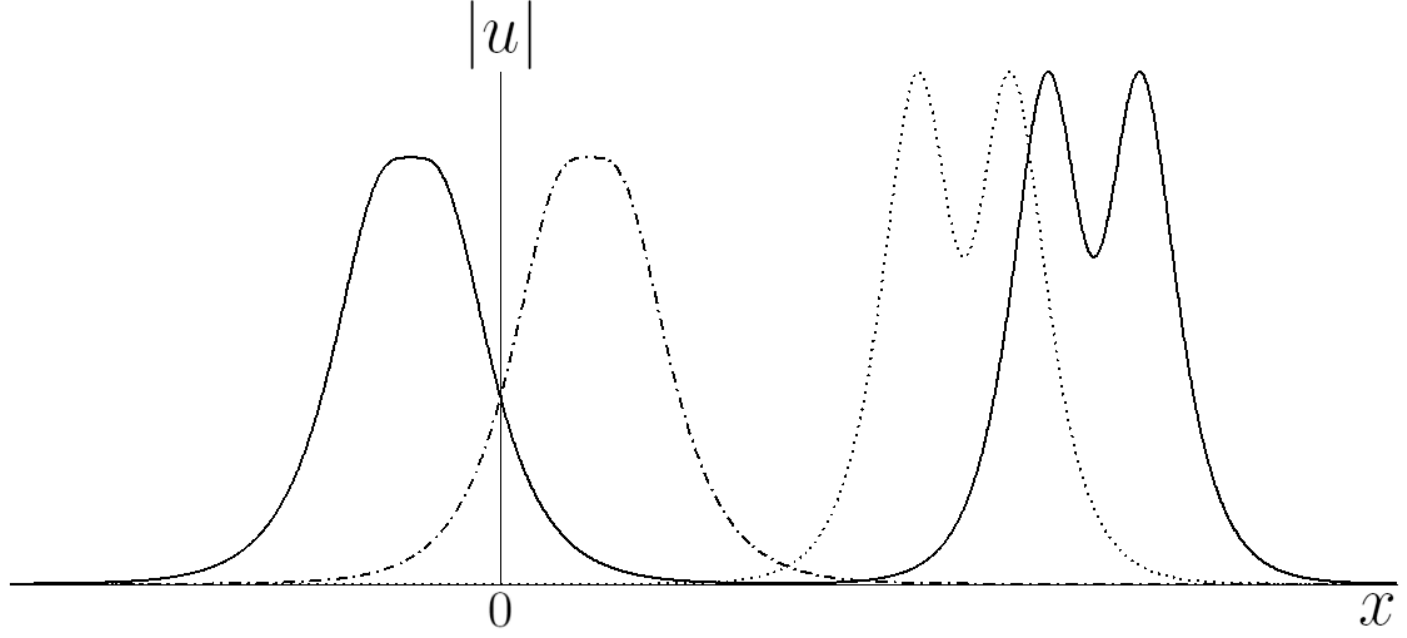}
\captionof{figure}{amplitude in asymptotic future}
\end{subfigure}%
\begin{subfigure}[t]{.5\textwidth}
\includegraphics[width=\textwidth]{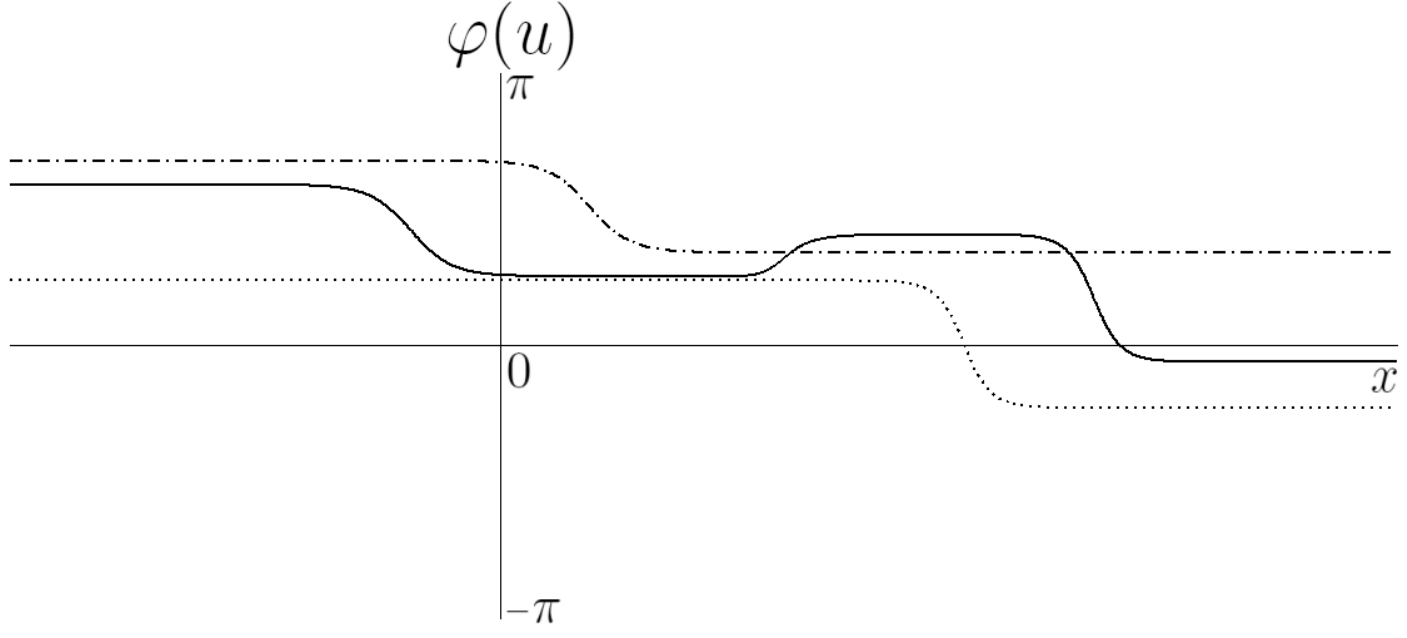}
\captionof{figure}{envelope phase in asymptotic future}
\end{subfigure}
\caption{Sasa-Satsuma collision (in solid) of 
a right-moving oscillatory soliton (in dots) and a standing-wave soliton (in dot-dash)
with $c_1=4$, $c_2=0$, $\nu_1=2$, $\nu_2=5$, $\varphi_1^-=0$, $\varphi_2^-=\pi/2$}
\label{standing_ss_right-overtake_shifts_c1=4_c2=0_nu1=2_nu2=5_phi1=0_phi2=halfpi}
\end{figure}

We remark that Theorems~\ref{thm:hshift} and~\ref{thm:ssshift} 
hold for collisions of an oscillatory wave and a standing wave. 
Thus, in a right-overtake collision with $c_1>c_2=0$,
the effect of the collision on the asymptotic positions of the waves 
is to shift the right-moving asymptotic oscillatory wave $u_1^\pm$ 
in a forward direction (\ie/ to the right, since $\Delta x_1>0$)
while the asymptotic standing wave $u_2^\pm$ is displaced 
in the opposite direction (\ie/ to the left, since $\Delta x_2<0$). 
Similarly, in a left-overtake collision with $0=c_1>c_2$,
the collision affects the asymptotic positions of the two waves 
by shifting the left-moving asymptotic oscillatory wave $u_2^\pm$ 
in a forward direction (\ie/ to the left, since $\Delta x_2<0$)
while the asymptotic standing wave $u_1^\pm$ is displayed 
in the opposite direction (\ie/ to the right, since $\Delta x_1>0$). 
In both cases, 
the asymptotic positions and the asymptotic phase angles of the waves 
are given by equations \eqref{asymptcomom}--\eqref{asymptphases}. 
The position and phase shifts are shown in 
\figref{standing_hirota_left-overtake_shifts_c1=0_c2=-4_nu1=2_nu2=5_phi1=0_phi2=halfpi}--\figref{standing_hirota_right-overtake_shifts_c1=4_c2=0_nu1=2_nu2=5_phi1=0_phi2=halfpi}
for the Hirota equation, 
and \figref{standing_ss_left-overtake_shifts_c1=0_c2=-4_nu1=2_nu2=5_phi1=0_phi2=halfpi}--\figref{standing_ss_right-overtake_shifts_c1=4_c2=0_nu1=2_nu2=5_phi1=0_phi2=halfpi} 
for the Sasa-Satsuma equation. 



\section{Interaction features and Concluding remarks}
\label{interactions} 

In previous work \cite{AncNgaWil}, 
collisions of ordinary solitary waves \eqref{solitarywave}
(\ie/ with no temporal harmonic modulation) have been studied
for the Hirota equation \eqref{hmkdveq} 
and the Sasa-Satsuma equation \eqref{ssmkdveq}. 
A collision in this case consists of 
a right-moving faster solitary wave 
with speed $c_1$ and phase $\phi_1$  
overtaking a right-moving slower solitary wave 
with speed $c_2$ and phase $\phi_2$. 
The corresponding ordinary $2$-soliton solutions 
exhibit several distinguishing properties. 
First, 
at a particular time $t=t_0$ the amplitude displays invariance 
$|u(t_0,x-\chi(t_0)|$ $=|u(t_0,\chi(t_0)-x|$ 
under spatial reflection around the center of momentum $x=\chi(t)$ of
the two solitary waves. 
This time $t=t_0$ can be understood to represent 
the moment of greatest nonlinear interaction of the waves during the collision. 
Second, the amplitude is always non-zero, $|u(t,x)|\neq 0$, 
throughout the collision.
As a consequence of these two properties, 
the interaction of the two waves can be characterized primarily by 
the convexity of $|u(t_0,x)|$ at the center of momentum $x=\chi(t_0)$
at time $t=t_0$. 
The case of negative convexity describes a collision 
such that the waves undergo a merge-split interaction
in which $|u(t_0,x)|$ has a single peak with an exponentially decreasing tail,
while the case of positive convexity describes 
a collision such that the waves exhibit either a bounce-exchange interaction 
in which $|u(t_0,x)|$ has a double peak with an exponentially decreasing tail,
or an absorb-emit interaction 
in which $|u(t_0,x)|$ has a pair of side peaks around a central peak
and an exponentially decreasing tail,
depending on the speed ratio and relative phase angle of the two waves, 
as explained in \Ref{AncNgaWil}. 

In contrast, 
collisions of oscillatory waves \eqref{oscil1soliton}
described by the $2$-soliton solutions from Proposition~\ref{thm:2soliton}
have very different features
(animations can be seen at 
http://lie.math.brocku.ca/\url{~}sanco/solitons/oscillatory.php):
\newline
(1) the amplitude $|u|$ exhibits invariance under spatial reflections only 
in special cases;
\newline
(2) the amplitude $|u|$ vanishes at certain positions $x$ and times $t$ 
(\ie/ $u$ has nodes);
\newline
(3) the phase $\arg(u)$ exhibits rapid spatial change at certain positions $x$ and times $t$
(\ie/ $u$ has phase coils with large spatial winding);
\newline
(4) the phase gradient $\arg(u)_x$ changes sign at certain positions $x$ and times $t$
(\ie/ $u$ has spatial reversals of phase winding). 

A detailed study of the interactions of oscillatory waves for these equations 
will be presented in a sequel paper. 
Our work in the present paper has two immediate extensions. 

First, 
the Hirota equation \eqref{hmkdveq} and the Sasa-Satsuma equation \eqref{ssmkdveq}
are known to be gauge-equivalent to third-order NLS equations \cite{GilHieNimOht}
\begin{equation}
q_{\tilde t} \pm i \sqrt{v/3}(3 q_{\tilde x\tilde x} + \alpha |q|^2 q)
+ \alpha |q|^2 q_{\tilde x} + \beta (|q|^2)_{\tilde x} q
+  q_{\tilde x\tilde x\tilde x} =0
\label{nls}
\end{equation}
through the Galilean-phase transformation
\EQ
\tilde t=t ,
\quad
\tilde x=x+vt ,
\qquad
u(t,x) = q(\tilde t,\tilde x)\exp\big(\pm i\sqrt{v/3}(\tilde x - (2v/3)\tilde t)\big)
\label{nlstransformation}
\endEQ
where $v>0$ is a speed parameter, 
with $\alpha=24$, $\beta=0$ in the Hirota case
and $\alpha=12$, $\beta=6$ in the Sasa-Satsuma case. 
Under this transformation, 
the oscillatory $1$-solitons \eqref{1soliton}--\eqref{ssoscilfunct}
shown in Proposition~\ref{thm:1soliton} for the Hirota and Sasa-Satsuma equations
correspond to NLS solitons of the same form
\begin{equation}
q(\tilde t,\tilde x) = \exp(i\phi)\exp(i\tilde\nu\tilde t)\tilde q(\tilde\xi), 
\quad
\tilde\xi = \tilde x-\tilde c\tilde t
\label{1solitonnls}
\end{equation}
parameterized by a speed $\tilde c$, a temporal frequency $\tilde \nu$, and a phase $\phi$, 
where
\begin{equation}
\tilde c = c+v ,
\quad
\tilde\nu = \nu\pm\sqrt{v/3}(c+4v/3) ,
\quad
\tilde q(\tilde\xi) = \exp(\mp i\sqrt{v/3}\;\tilde\xi)\tilde f(\tilde\xi) . 
\end{equation}
Consequently, 
we obtain NLS oscillatory $2$-solitons 
\begin{equation}
q(\tilde t,\tilde x) = 
\exp(i\phi_1)\exp(i\tilde\nu_1\tilde t)\tilde q_1(\tilde\xi_1,\tilde\xi_2)
+ \exp(i\phi_2)\exp(i\tilde\nu_2\tilde t)\tilde q_2(\tilde\xi_1,\tilde\xi_2)
\label{2solitonnls}
\end{equation}
with
\begin{gather}
\tilde\xi_1 = \tilde x-\tilde c_1\tilde t ,
\quad
\tilde\xi_2 = \tilde x-\tilde c_2\tilde t , 
\\
\tilde c_1 = c_1+v ,
\quad
\tilde c_2 = c_2+v , 
\\
\tilde\nu_1 = \nu_1\pm\sqrt{v/3}(c_1+4v/3) ,
\quad
\tilde\nu_2 = \nu_2\pm\sqrt{v/3}(c_2+4v/3) ,
\\
\tilde q_1(\tilde\xi_1,\tilde\xi_2) 
= \exp(\mp i\sqrt{v/3}\;\tilde\xi_1)\tilde f_1(\tilde\xi_1,\tilde\xi_2) ,
\quad
\tilde q_2(\tilde\xi_1,\tilde\xi_2) 
= \exp(\mp i\sqrt{v/3}\;\tilde\xi_2)\tilde f_2(\tilde\xi_1,\tilde\xi_2) ,
\end{gather}
where the functions 
$\tilde f_1(\tilde\xi_1,\tilde\xi_2)$ and $\tilde f_2(\tilde\xi_1,\tilde\xi_2)$
are given in Proposition~\ref{thm:2soliton} 
for the oscillatory $2$-solitons \eqref{2soliton}--\eqref{ssoscilY}
of the Hirota and Sasa-Satsuma equations. 
In addition, 
we obtain NLS oscillatory breathers in the special case $c_1=c_2\neq 0$,
discussed in \Ref{AncNgaWil}. 

The main results stated in 
Theorems~\ref{thm:hshift} and~\ref{thm:ssshift}
on the properties of collisions described by oscillatory $2$-solitons,
carry over directly to the third-order NLS equation \eqref{nls}.
In particular, 
the net effect of a collision is to shift 
the asymptotic positions and phases of the individual oscillatory waves 
while the speed and the temporal frequency of each wave remains unchanged,
such that the center of momentum of the waves is preserved in the collision. 

Second, 
the Hirota equation \eqref{hmkdveq} has two natural multi-component
generalizations given by $U(N)$-invariant integrable mKdV equations \cite{Anc}
\EQ
\vec u_t +12( |\vec u|^2\vec u_x +(\vec u_x\cdot\overline{\vec u})\vec u )+ \vec u_{xxx}=0
\label{vecmkdv1}
\endEQ
and 
\EQ
\vec u_t+24( |\vec u|^2\vec u_x +(\vec u_x\cdot\overline{\vec u})\vec u -(\vec u_x\cdot\vec u)\overline{\vec u} ) +\vec u_{xxx}=0
\label{vecmkdv2}
\endEQ
where $\vec u(t,x)$ is a $N$-component complex vector variable. 
For all $N\geq 2$, these two vector equations 
admit vector oscillatory wave solutions of the form 
\EQ
\vec u(t,x) = \exp(i\nu t)\tilde f_{\rm H}(x-ct)\hat\psi 
\endEQ
with wave speed $c$ and temporal frequency $\nu$, 
satisfying the kinematic relation \eqref{1solitonkinrel}, 
where $\hat\psi$ is an arbitrary constant complex unit vector
and $\tilde f_{\rm H}$ is the complex envelope function \eqref{hoscilfunct}
for the oscillatory $1$-soliton solution of 
the scalar Hirota equation \eqref{hmkdveq}. 
In forthcoming work, we plan to generalize the results in the present paper 
to study the vector oscillatory $2$-soliton solutions 
and vector oscillatory breather solutions of 
both equations \eqref{vecmkdv1} and \eqref{vecmkdv2}.

\section*{Acknowledgement}
S. Anco is supported by an NSERC research grant. 
The authors thank Nestor Tchegoum Ngatat for assistance 
in an early stage of this work. 

Email:
\lowercase{\scshape{sanco@brocku.ca}},
\lowercase{\scshape{sattar\_ju@yahoo.com}},
\lowercase{\scshape{markw@math.ubc.ca}}

\end{document}